\pdfoutput=1
\documentclass[12pt,a4paper]{article}
\usepackage{ifthen}
\usepackage{comment}
\usepackage{graphicx}
\usepackage{float}
\newboolean{pdflatex}
\setboolean{pdflatex}{true} 
\newboolean{articletitles}
\setboolean{articletitles}{true}
\newboolean{uprightparticles}
\setboolean{uprightparticles}{false}
\def\paperauthors{P.~Alvarez~Cartelle, A.~Anelli, A.~Balboni, A.~Beck, F.~Betti, J.~Bex, F.~Borgato, R.~Calabrese, M.~Calvi, R.~Cardinale, G.~Cavallero, L.~Cojocariu, D.C.~Craik, E.~Dall'Occo, C.~D'Ambrosio, V.~Duk, L.~Fantini, F.~Ferrari, M.~Fiorini, E.~Franzoso, C.~Frei, S.~Gambetta, S.~Ghizzo, Z.~Ghorbanimoghaddam, C.~Gotti, T.~Gys, C.R.~Jones, I.~Juszczak, M.~Kane, F.~Keizer, M.~Kucharczyk, T.~Long, F.~Maciuc, L.~Malentacca, G.~Martelli, A.~Mauri, M.~McCann, F.~Muheim, E.M.~Niel, S.~Okamura, F.~Oliva, A.~Papanestis, M.~Patel, J.~Patoc, M.~Piccini, V.~Placinta, J.H.~Rademacker, J.~Reich, G.~Romolini, A.~Sergi, G.~Simi, M.~Smith, F.~Swystun, H.~Tilquin, C.~Vrahas, G.~Wilkinson, R.~Williams, M.~Witek, S.A.~Wotton, S.~Xian, Y.~Yang}
\def\paperasciititle{Luminosity measurement with the LHCb RICH detectors in Run 3} 
\def\papertitle{Luminosity measurement with the \lhcb \rich detectors in Run 3}
\def\paperkeywords{{Photon detectors for UV, visible and IR photons (vacuum) (photomultipliers, HPDs, others)}, {Cherenkov detectors}, {Performance of High Energy Physics Detectors}, {Detector modelling and simulations II (electric fields, charge transport, multiplication and induction, pulse formation, electron emission, etc)}} 
\def\papercopyright{\the\year\ CERN for the benefit of the LHCb collaboration}
\def\paperlicence{CC BY 4.0 licence}
\def\paperlicenceurl{https://creativecommons.org/licenses/by/4.0/}

\usepackage[top=1in, bottom=1.25in, left=1in, right=1in]{geometry}

\usepackage{microtype}
\usepackage{lineno}  
\usepackage{xspace} 
\usepackage{caption} 

\usepackage{graphicx}  
\usepackage{color}
\usepackage{colortbl}
\graphicspath{{./figs/}} 

\usepackage{amsmath} 
\usepackage{amssymb}
\usepackage{amsfonts}
\usepackage{upgreek} 

\newcommand*\patchAmsMathEnvironmentForLineno[1]{%
\expandafter\let\csname old#1\expandafter\endcsname\csname #1\endcsname
\expandafter\let\csname oldend#1\expandafter\endcsname\csname
end#1\endcsname
 \renewenvironment{#1}%
   {\linenomath\csname old#1\endcsname}%
   {\csname oldend#1\endcsname\endlinenomath}%
}
\newcommand*\patchBothAmsMathEnvironmentsForLineno[1]{%
  \patchAmsMathEnvironmentForLineno{#1}%
  \patchAmsMathEnvironmentForLineno{#1*}%
}
\AtBeginDocument{%
\patchBothAmsMathEnvironmentsForLineno{equation}%
\patchBothAmsMathEnvironmentsForLineno{align}%
\patchBothAmsMathEnvironmentsForLineno{flalign}%
\patchBothAmsMathEnvironmentsForLineno{alignat}%
\patchBothAmsMathEnvironmentsForLineno{gather}%
\patchBothAmsMathEnvironmentsForLineno{multline}%
\patchBothAmsMathEnvironmentsForLineno{eqnarray}%
}

\usepackage{hyperxmp}

\usepackage[pdftex,
            pdfauthor={\paperauthors},
            pdftitle={\paperasciititle},
            pdfkeywords={\paperkeywords},
            pdfcopyright={Copyright (C) \papercopyright},
            pdflicenseurl={\paperlicenceurl}]{hyperref}

\usepackage[colorinlistoftodos,textsize=scriptsize]{todonotes}

\usepackage[bottom,flushmargin,hang,multiple]{footmisc}

\usepackage[all]{hypcap} 

\usepackage{xspace} 
\usepackage{upgreek}

\def\lhcb   {\mbox{LHCb}\xspace}

\def\lhc    {\mbox{LHC}\xspace}

\def\velo   {VELO\xspace}
\def\rich   {RICH\xspace}
\def\richone {RICH1\xspace}
\def\richtwo {RICH2\xspace}

\def\MagUp {\mbox{\em Mag\kern -0.05em Up}\xspace}

\def\ecs    {ECS\xspace}

\ifthenelse{\boolean{uprightparticles}}%
{

 \def\PDelta      {\ensuremath{\Delta}\xspace}                 
 \def\PXi         {\ensuremath{\Xi}\xspace}                 
 \def\PLambda     {\ensuremath{\Lambda}\xspace}                 
 \def\PSigma      {\ensuremath{\Sigma}\xspace}                 
 \def\POmega      {\ensuremath{\Omega}\xspace}                 
 \def\PUpsilon    {\ensuremath{\Upsilon}\xspace}
 \let\oldPi\Pi
 \def\PPi         {\ensuremath{\oldPi}\xspace}

 \def\PB      {\ensuremath{\mathrm{B}}\xspace}                 
                  
 \def\PD      {\ensuremath{\mathrm{D}}\xspace}

 \def\PK      {\ensuremath{\mathrm{K}}\xspace}

 \def\Pi      {\ensuremath{\mathrm{i}}\xspace}

 \def\Ps      {\ensuremath{\mathrm{s}}\xspace}

 \def\thebaroffset{0.0em}
}
{

 \mathchardef\PDelta="7101
 \mathchardef\PXi="7104
 \mathchardef\PLambda="7103
 \mathchardef\PSigma="7106
 \mathchardef\POmega="710A
 \mathchardef\PUpsilon="7107
 \mathchardef\PPi="7105
                  
 \def\PB      {\ensuremath{B}\xspace}                 
                  
 \def\PD      {\ensuremath{D}\xspace}

 \def\PK      {\ensuremath{K}\xspace}

 \def\Pi      {\ensuremath{i}\xspace}

 \def\Ps      {\ensuremath{s}\xspace}

 \def\thebaroffset{0.18em}
}
\newcommand{\offsetoverline}[2][\thebaroffset]{\kern #1\overline{\kern -#1 #2}}%

\makeatletter
\ifcase \@ptsize \relax
  \newcommand{\miniscule}{\@setfontsize\miniscule{4}{5}}
\or
  \newcommand{\miniscule}{\@setfontsize\miniscule{5}{6}}
\or
  \newcommand{\miniscule}{\@setfontsize\miniscule{5}{6}}
\fi
\makeatother

\DeclareRobustCommand{\optbar}[1]{\shortstack{{\miniscule (\rule[.5ex]{1.25em}{.18mm})}
  \\ [-.7ex] $#1$}}

\def\squark    {{\ensuremath{\Ps}}\xspace}

\def\KorKbar {\kern \thebaroffset\optbar{\kern -\thebaroffset \PK}{}\xspace}

\def\D       {{\ensuremath{\PD}}\xspace}

\def\DorDbar {\kern \thebaroffset\optbar{\kern -\thebaroffset \PD}\xspace}

\def\Dp      {{\ensuremath{\D^+}}\xspace}
\def\Dm      {{\ensuremath{\D^-}}\xspace}

\def\DpDm    {\ensuremath{\Dp {\kern -0.16em \Dm}}\xspace}

\def\B       {{\ensuremath{\PB}}\xspace}

\def\BorBbar {\kern \thebaroffset\optbar{\kern -\thebaroffset \PB}\xspace}

\def\Bd      {{\ensuremath{\B^0}}\xspace}

\def\BdorBdbar {\kern \thebaroffset\optbar{\kern -\thebaroffset \Bd}\xspace}

\def\Bs      {{\ensuremath{\B^0_\squark}}\xspace}

\def\BsorBsbar {\kern \thebaroffset\optbar{\kern -\thebaroffset \Bs}\xspace}

\def\Y#1S{\ensuremath{\PUpsilon{(#1S)}}\xspace}

\def\LorLbar     {\kern \thebaroffset\optbar{\kern -\thebaroffset \PLambda}\xspace}

\def\order   {{\ensuremath{\mathcal{O}}}\xspace}

\def\AT#1     {\ensuremath{A_{\mathrm{T}}^{#1}}\xspace}           

\def\C#1      {\ensuremath{\mathcal{C}_{#1}}\xspace}                       
\def\Cp#1     {\ensuremath{\mathcal{C}_{#1}^{'}}\xspace}                    
\def\Ceff#1   {\ensuremath{\mathcal{C}_{#1}^{\mathrm{(eff)}}}\xspace}        
\def\Cpeff#1  {\ensuremath{\mathcal{C}_{#1}^{'\mathrm{(eff)}}}\xspace}       
\def\Ope#1    {\ensuremath{\mathcal{O}_{#1}}\xspace}                       
\def\Opep#1   {\ensuremath{\mathcal{O}_{#1}^{'}}\xspace}                    

\newcommand{\nospaceunit}[1]{\ensuremath{\text{#1}}}       
\newcommand{\aunit}[1]{\ensuremath{\text{\,#1}}}       

\newcommand{\tev}{\aunit{Te\kern -0.1em V}\xspace}
\newcommand{\gev}{\aunit{Ge\kern -0.1em V}\xspace}
\newcommand{\mev}{\aunit{Me\kern -0.1em V}\xspace}
\newcommand{\kev}{\aunit{ke\kern -0.1em V}\xspace}
\newcommand{\ev}{\aunit{e\kern -0.1em V}\xspace}
 
\newcommand{\mevc}{\ensuremath{\aunit{Me\kern -0.1em V\!/}c}\xspace}
\newcommand{\gevc}{\ensuremath{\aunit{Ge\kern -0.1em V\!/}c}\xspace}
\newcommand{\mevcc}{\ensuremath{\aunit{Me\kern -0.1em V\!/}c^2}\xspace}
\newcommand{\gevcc}{\ensuremath{\aunit{Ge\kern -0.1em V\!/}c^2}\xspace}

\def\cma  {\ensuremath{\aunit{cm}^2}\xspace}

\def\mbarn{\aunit{mb}\xspace}
\def\mub{\ensuremath{\,\upmu\nospaceunit{b}}\xspace}

\def\mus  {\ensuremath{\,\upmu\nospaceunit{s}}\xspace}

\def\mhz  {\ensuremath{\aunit{MHz}}\xspace}
\def\khz  {\ensuremath{\aunit{kHz}}\xspace}
\def\hz   {\ensuremath{\aunit{Hz}}\xspace}

\def\order{{\ensuremath{\mathcal{O}}}\xspace}

\def\gsim{{~\raise.15em\hbox{$>$}\kern-.85em
          \lower.35em\hbox{$\sim$}~}\xspace}
\def\lsim{{~\raise.15em\hbox{$<$}\kern-.85em
          \lower.35em\hbox{$\sim$}~}\xspace}

\newcommand{\lum} {\ensuremath{\mathcal{L}}\xspace}

\def\tell1  {TELL1\xspace}
\def\ukl1   {UKL1\xspace}

\newcommand{\eg}{\mbox{\itshape e.g.}\xspace}
\newcommand{\ie}{\mbox{\itshape i.e.}\xspace}

\newcommand{\lhcborcid}[1]{\href{https://orcid.org/#1}{\hspace*{0.1em}\raisebox{-0.45ex}{\includegraphics[width=1em]{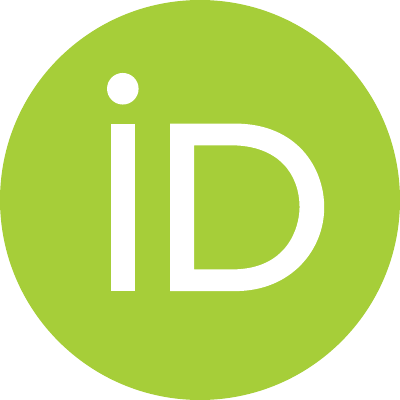}}}}

\usepackage{cite} 
\usepackage{mciteplus}

\usepackage{longtable} 
\usepackage{subfigure}
\usepackage{subcaption}

\begin{document}

\renewcommand{\thefootnote}{\fnsymbol{footnote}}
\setcounter{footnote}{1}

\begin{titlepage}
\pagenumbering{roman}

\vspace*{-1.5cm}
\centerline{\large EUROPEAN ORGANIZATION FOR NUCLEAR RESEARCH (CERN)}
\vspace*{1.5cm}
\noindent
\begin{tabular*}{\linewidth}{lc@{\extracolsep{\fill}}r@{\extracolsep{0pt}}}
\ifthenelse{\boolean{pdflatex}}
{\vspace*{-1.5cm}\mbox{\!\!\!\includegraphics[width=.14\textwidth]{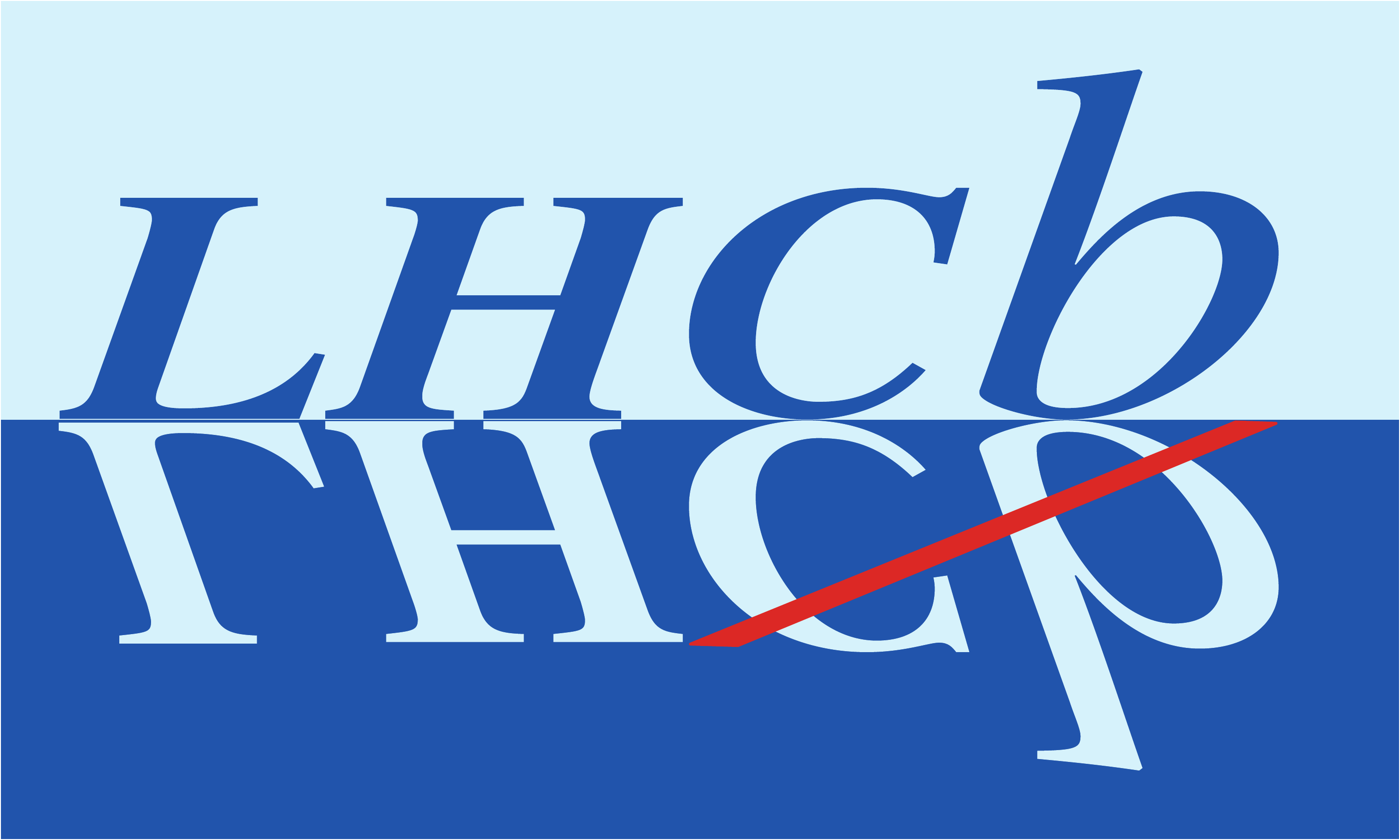}} & &} 
{\vspace*{-1.2cm}\mbox{\!\!\!\includegraphics[width=.12\textwidth]{lhcb-logo.eps}} & &}
\\
 & & CERN-LHCb-DP-2024-003 \\  
 & & \today \\
 & & \\
\end{tabular*}

\vspace*{1.0cm} 

{\normalfont\bfseries\boldmath\huge
\begin{center}
  \papertitle 
\end{center}
}

\vspace*{0.5cm} 

\begin{center}
\centerline
\small
P.~Alvarez~Cartelle$^{13}$\lhcborcid{0000-0003-1652-2834},
A.~Anelli$^{5,d,10}$\lhcborcid{0000-0002-6191-934X},
A.~Balboni$^{3}$\lhcborcid{0009-0003-8872-976X},
A.~Beck$^{18}$\lhcborcid{0000-0003-4872-1213},
F.~Betti$^{15}$\lhcborcid{0000-0002-2395-235X},
J.~Bex$^{13}$\lhcborcid{0000-0002-2856-8074},
F.~Borgato$^{6,10}$\lhcborcid{0000-0002-3149-6710},
R.~Calabrese$^{3,b}$\lhcborcid{0000-0002-1354-5400},
M.~Calvi$^{5,d}$\lhcborcid{0000-0002-8797-1357},
R.~Cardinale$^{4,c}$\lhcborcid{0000-0002-7835-7638},
G.~Cavallero$^{3}$\lhcborcid{0000-0002-8342-7047},
L.~Cojocariu$^{9}$\lhcborcid{0000-0002-1281-5923},
D.C.~Craik$^{11}$\lhcborcid{0000-0002-3684-1560},
E.~Dall'Occo$^{10}$\lhcborcid{0000-0001-9313-4021},
C.~D'Ambrosio$^{16}$\lhcborcid{0000-0003-4344-9994},
V.~Duk$^{7}$\lhcborcid{0000-0001-6440-0087},
L.~Fantini$^{7,f,10}$\lhcborcid{0000-0002-2351-3998},
F.~Ferrari$^{2,a}$\lhcborcid{0000-0002-3721-4585},
M.~Fiorini$^{3,b}$\lhcborcid{0000-0001-6559-2084},
E.~Franzoso$^{3,b,*}$\lhcborcid{0000-0003-2130-1593},
C.~Frei$^{10}$\lhcborcid{0000-0001-5501-5611},
S.~Gambetta$^{15}$\lhcborcid{0000-0003-2420-0501},
S.~Ghizzo$^{4,c}$\lhcborcid{0009-0001-5178-9385},
Z.~Ghorbanimoghaddam$^{12}$\lhcborcid{0000-0002-4410-9505},
C.~Gotti$^{5}$\lhcborcid{0000-0003-2501-9608},
T.~Gys$^{10}$\lhcborcid{0000-0002-6825-6497},
C.R.~Jones$^{13}$\lhcborcid{0000-0003-1699-8816},
I.~Juszczak$^{8}$\lhcborcid{0000-0002-1285-3911},
M.~Kane$^{15}$\lhcborcid{0009-0006-5064-966X},
F.~Keizer$^{10}$\lhcborcid{0000-0002-1290-6737},
M.~Kucharczyk$^{8}$\lhcborcid{0000-0003-4688-0050},
T.~Long$^{13}$\lhcborcid{0000-0001-7292-848X},
F.~Maciuc$^{9}$\lhcborcid{0000-0001-6651-9436},
L.~Malentacca$^{10}$\lhcborcid{0000-0001-6717-2980},
G.~Martelli$^{7}$\lhcborcid{0000-0002-6150-3168},
A.~Mauri$^{16}$\lhcborcid{0000-0003-1664-8963},
M.~McCann$^{16}$\lhcborcid{0000-0002-3038-7301},
F.~Muheim$^{15}$\lhcborcid{0000-0002-1131-8909},
E.M.~Niel$^{1}$\lhcborcid{0000-0002-6587-4695},
S.~Okamura$^{3,b}$\lhcborcid{0000-0003-1229-3093},
F.~Oliva$^{15}$\lhcborcid{0000-0001-7025-3407},
A.~Papanestis$^{14,10}$\lhcborcid{0000-0002-5405-2901},
M.~Patel$^{16}$\lhcborcid{0000-0003-3871-5602},
J.~Patoc$^{17}$\lhcborcid{0009-0000-1201-4918},
M.~Piccini$^{7}$\lhcborcid{0000-0001-8659-4409},
V.~Placinta$^{9}$\lhcborcid{0000-0003-4465-2441},
J.H.~Rademacker$^{12}$\lhcborcid{0000-0003-2599-7209},
J.~Reich$^{12}$\lhcborcid{0000-0002-2657-4040},
G.~Romolini$^{3,10}$\lhcborcid{0000-0002-0118-4214},
A.~Sergi$^{4,c}$\lhcborcid{0000-0001-9495-6115},
G.~Simi$^{6,e}$\lhcborcid{0000-0001-6741-6199},
M.~Smith$^{16}$\lhcborcid{0000-0002-3872-1917},
F.~Swystun$^{13}$\lhcborcid{0009-0006-0672-7771},
H.~Tilquin$^{16}$\lhcborcid{0000-0003-4735-2014},
C.~Vrahas$^{15}$\lhcborcid{0000-0001-6104-1496},
G.~Wilkinson$^{17,10}$\lhcborcid{0000-0001-5255-0619},
R.~Williams$^{13}$\lhcborcid{0000-0002-2675-3567},
M.~Witek$^{8}$\lhcborcid{0000-0002-8317-385X},
S.A.~Wotton$^{13}$\lhcborcid{0000-0003-4543-8121},
S.~Xian$^{19}$\lhcborcid{0009-0009-9115-1122},
Y.~Yang$^{4,c}$\lhcborcid{0000-0002-8917-2620}.

\bigskip
\noindent
{\footnotesize \it
$^{1}$Laboratoire Leprince-Ringuet, CNRS/IN2P3, Ecole Polytechnique, Institut Polytechnique de Paris, Palaiseau, France\\
$^{2}$INFN Sezione di Bologna, Bologna, Italy\\
$^{3}$INFN Sezione di Ferrara, Ferrara, Italy\\
$^{4}$INFN Sezione di Genova, Genova, Italy\\
$^{5}$INFN Sezione di Milano-Bicocca, Milano, Italy\\
$^{6}$INFN Sezione di Padova, Padova, Italy\\
$^{7}$INFN Sezione di Perugia, Perugia, Italy\\
$^{8}$Henryk Niewodniczanski Institute of Nuclear Physics  Polish
Academy of Sciences, Krak{\'o}w, Poland\\
$^{9}$Horia Hulubei National Institute of Physics and Nuclear Engineering, Bucharest-Magurele, Romania\\
$^{10}$European Organization for Nuclear Research (CERN), Geneva,
Switzerland\\
$^{11}$Physik-Institut, Universit{\"a}t Z{\"u}rich, Z{\"u}rich,
Switzerland\\
$^{12}$H.H. Wills Physics Laboratory, University of Bristol, Bristol, United Kingdom\\
$^{13}$Cavendish Laboratory, University of Cambridge, Cambridge,
United Kingdom\\
$^{14}$STFC Rutherford Appleton Laboratory, Didcot, United Kingdom\\
$^{15}$School of Physics and Astronomy, University of Edinburgh,
Edinburgh, United Kingdom\\
$^{16}$Imperial College London, London, United Kingdom\\
$^{17}$Department of Physics, University of Oxford, Oxford, United Kingdom\\
$^{18}$Massachusetts Institute of Technology, Cambridge, MA, United
States\\
$^{19}$Guangdong Provincial Key Laboratory of Nuclear Science, Guangdong-Hong Kong Joint Laboratory of Quantum Matter, Institute of Quantum Matter, South China Normal University, Guangzhou, China\\
\bigskip
\noindent
$^{a}$Universit{\`a} di Bologna, Bologna, Italy\\
$^{b}$Universit{\`a} di Ferrara, Ferrara, Italy\\
$^{c}$Universit{\`a} di Genova, Genova, Italy\\
$^{d}$Universit{\`a} degli Studi di Milano-Bicocca, Milano, Italy\\
$^{e}$Universit{\`a} di Padova, Padova, Italy\\
$^{f}$Universit{\`a}  di Perugia, Perugia, Italy\\
\bigskip
\noindent
$^*$Corresponding author\\
}

\end{center}

\vspace{\fill}

\begin{abstract}
  \noindent
  The \lhcb Ring-Imaging Cherenkov detectors are built to provide charged hadron identification over a large range of momentum. The upgraded detectors are also capable of providing an independent measurement of the luminosity for the LHCb experiment during LHC Run 3. The modelling of the opto-electronics chain, the application of the powering strategy during operations, the calibration procedures and the proof of principle of a novel technique for luminosity determination are presented. In addition, the preliminary precision achieved during the 2023 data-taking year for real-time and offline luminosity measurements is reported. 
  
\end{abstract}

\vspace*{2.0cm}

\begin{center}
  Submitted to JINST
\end{center}

\vspace{\fill}

{\footnotesize 
\centerline{\copyright~\papercopyright. \href{\paperlicenceurl}{\paperlicence}.}}
\vspace*{2mm}

\end{titlepage}

\newpage
\setcounter{page}{3}
\mbox{~}

\renewcommand{\thefootnote}{\arabic{footnote}}
\setcounter{footnote}{0}

\cleardoublepage

\pagestyle{plain} 
\setcounter{page}{1}
\pagenumbering{arabic}

\section{Introduction}
\label{sec:Introduction}

An essential component of the \lhcb flavour physics programme is the discrimination of charged hadrons to distinguish between pions, kaons and protons. The Ring-Imaging Cherenkov (\rich) detectors provide charged hadron identification within the momentum range of approximately 3 to 100\gevc and play a crucial role in \lhcb measurements using data from \lhc Run 1 and 2, as detailed in Ref.~\cite{LHCb-DP-2012-003,Calabrese:2022eju}.

The \rich system comprises two detectors, namely \richone and \richtwo, employing a system of spherical and planar mirrors to focus Cherenkov photons, generated by charged tracks within fluorocarbon gaseous radiators, onto the photon detection planes. \richone covers the complete spectrometer angular acceptance from 25 to 300 mrad and is positioned upstream the \lhcb dipole magnet. \richone measures the low-to-intermediate momentum range of 3 - 40\gevc. The angular acceptance of the \richtwo detector, located downstream the \lhcb dipole, is 15 - 120 mrad and is optimised to cover the high momentum range 15 - 100\gevc.

\begin{figure}[tb]
  \begin{center}
    \includegraphics[width=0.49\linewidth]{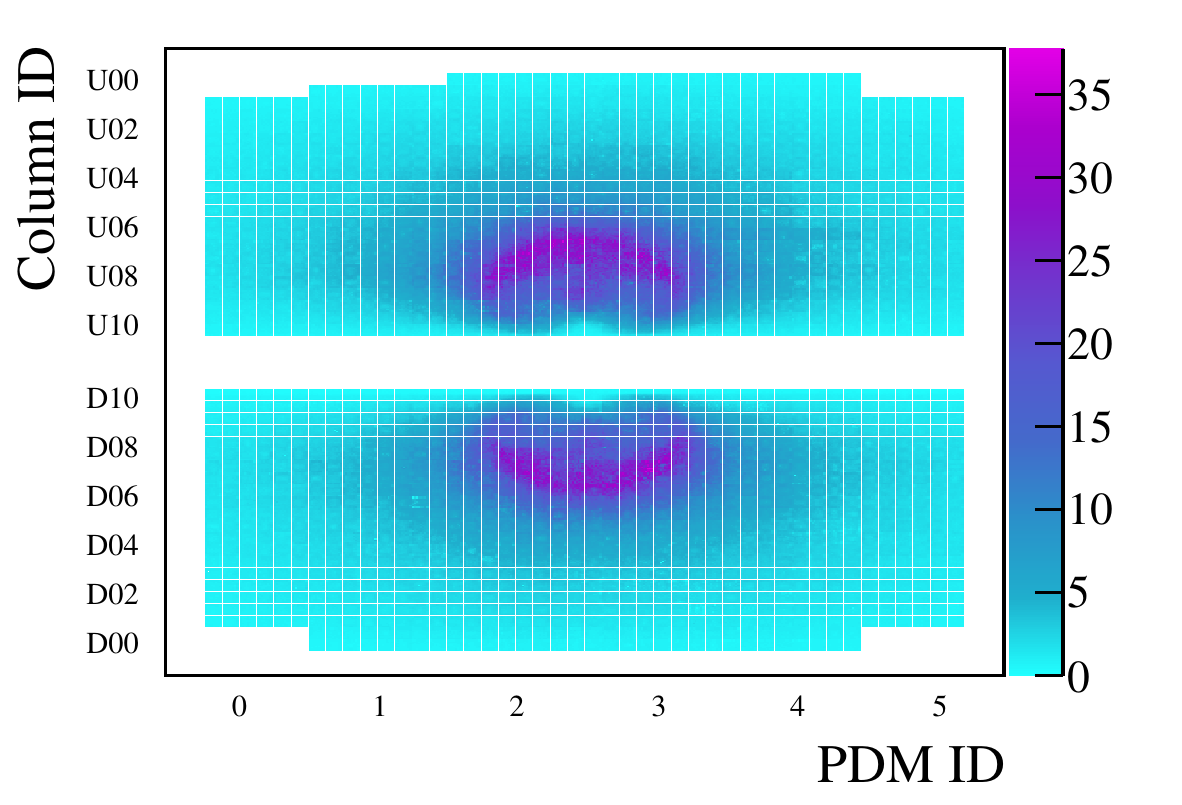}
    \includegraphics[width=0.49\linewidth]{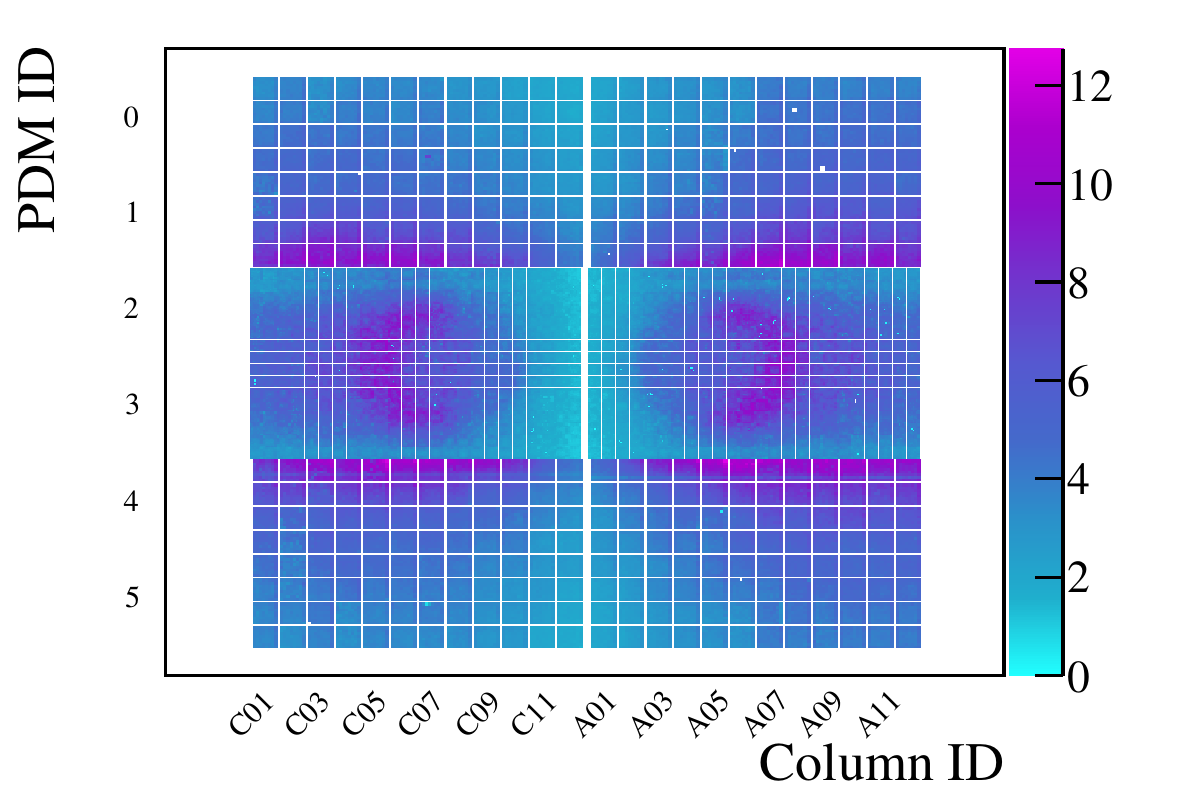}
    \vspace*{-0.5cm}
  \end{center}
  \caption{Two-dimensional hit-maps in percent for \richone (left) and \richtwo (right). The events are required to meet the first level of physics selections, \ie the fraction of fired pixels is larger than the fraction for minimum-bias events.}
  \label{fig:hitmaps}
\end{figure}

To efficiently take advantage of the five-fold increase in the instantaneous luminosity, reaching \lum = $2 \times 10^{33}~\mathrm{cm}^{-2}~\mathrm{s}^{-1}$ in \lhc Run 3, the \lhcb detector is read out at the full \lhc bunch crossing rate of 40\mhz. The \rich system, previously employing Hybrid Photon Detectors (HPDs)~\cite{TA2:1999awj} with embedded front-end electronics read out at 1\mhz, has been upgraded. The HPDs have subsequently been replaced with 64 channels Multianode Photomultiplier Tubes (MaPMTs)~\cite{Cadamuro:2014hza,Calvi:2015yra} from Hamamatsu Photonics (HPK) and new front-end electronics transmitting data at 40\mhz. The MaPMTs employed by the \rich system are the 1-inch HPK-R11265-M64 in \richone and the inner region of \richtwo, and the 2-inch HPK-R12699-M64 in the outer region of \richtwo. Their operating parameters are tailored to detect single photons over a wide range of illumination rates. Detection rates vary from approximately 100\mhz/\cma in the central area of \richone, to lower rates of about 5\mhz/\cma in the peripheral region of \richtwo. Sixteen (four) HPK-R11265-M64 (HPK-R12699-M64) MaPMTs are grouped into a unit called Photon Detector Module (PDM). Six PDMs build a so-called column, for a total of 22 columns oriented along the $x$ coordinate in \richone and 24 columns developing along the $y$ coordinate in \richtwo, where $z$ is the direction of the beams. An example of two-dimensional hit-maps are shown in Fig.~\ref{fig:hitmaps}. The new \rich detectors are described in detail in Ref.~\cite{LHCb:2023hlw} and the preliminary figures of merit for hadron identification with early \lhc Run 3 data are reported in Ref.~\cite{FRANZOSO2024169689}, showing already an excellent performance as by design.

In the \lhcb experiment, an automated luminosity levelling procedure is adopted to limit the instantaneous luminosity to \lum = $2 \times 10^{33}~\mathrm{cm}^{-2}~\mathrm{s}^{-1}$ and to compensate the luminosity decay during a fill. This strategy allows to optimise the vertexing, tracking and particle-identification performance by reducing the systematic uncertainties associated to detector occupancy changes. The levelling is implemented by the \lhc machine by changing the overlap of the beams with the so-called "separation levelling" in steps, to keep the instantaneous luminosity within $\pm 5\%$ with respect to the target requested by \lhcb~\cite{Alemany-Fernandez:2013bya}. The instantaneous luminosity is measured in \lhcb and provided to the \lhc by the Probe for Luminosity Measurement (PLUME) system~\cite{LHCb-TDR-022}.   

In this paper, the use of the \rich system to provide a standalone and alternative measurement of the \lhcb luminosity is reported. Two novel measurement techniques are used: the first, providing a real-time measurement of the instantaneous luminosity, takes advantage of the determination of the MaPMT anode currents; the second employs the Cherenkov hits as digitised after the opto-electronics chain, made possible by the low noise (order \khz/\cma) of the \rich detectors. An additional, independent system, designed to act as a safety measure for environmental light-leak detection inside the MaPMT enclosures, is used to provide further online measurements of the instantaneous luminosity.

In the \richone detector, the MaPMTs are operated by using a powering scheme involving the bias of the last of the twelve dynodes, in order to maintain a linear response of the MaPMTs with the expected rates, as described in Sec.~\ref{sec:spice}. The implementation of the method is described in Sec.~\ref{sec:operations}, where it is shown how an output signal proportional to the luminosity is obtained through the measurement of the power-supply currents. The possibility to use the number of Cherenkov hits for the same purpose is verified through simulated samples generated with different pile-up conditions, as described in Sec.~\ref{sec:sim}, and calibrated through van der Meer scans performed during the 2023 data-taking period as shown in Sec.~\ref{sec:VdM}. 
To calibrate the MaPMT anode currents for providing a real-time luminosity signal, a dedicated data-taking phase is performed in which the instantaneous luminosity requested from the \lhc is varied in known steps, as described in Sec.~\ref{sec:currents}. Finally, the results and future prospects are presented in Sec.~\ref{sec:results}.

\section{Modelling of the MaPMT response}
\label{sec:spice}

\begin{figure}[tb]
	\centering
	\includegraphics[width=\textwidth]{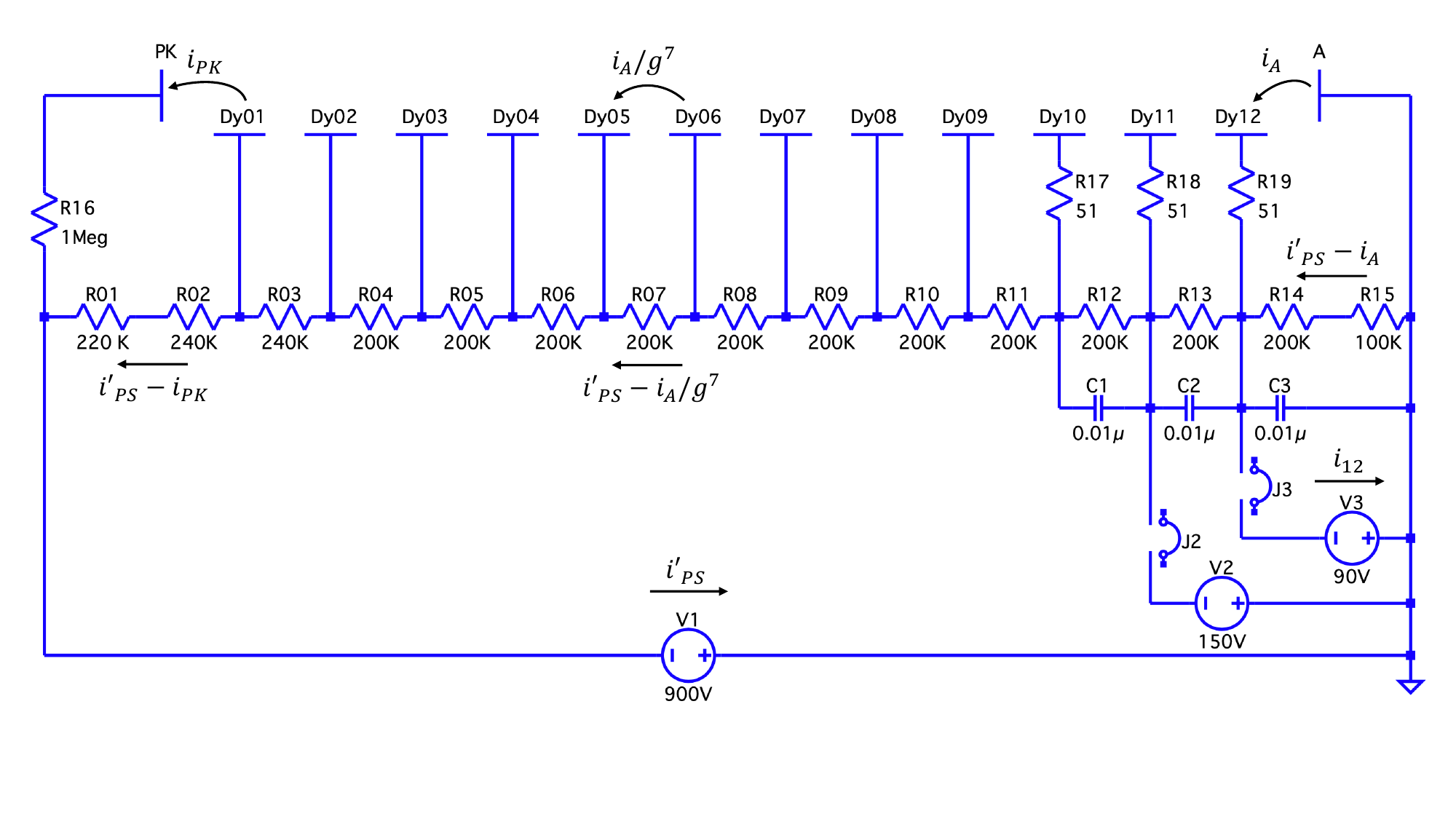}
	\caption{Electrical scheme of the HV voltage divider for tube HPK-R11265-M64, used in \richone and in the inner region of \richtwo, with three power supplies. The currents flowing through the voltage divider resistors are only shown for the first, sixth and last stages. The optional additional power supply lines are shown, corresponding to voltages $V_\mathrm{2}$ and $V_\mathrm{3}$.}
	\label{fig:VD_EC-R}
\end{figure}

For tube HPK-R11265-M64, equipped with 12 dynodes, the standard voltage divider suggested by the manufacturer, shown in Fig.~\ref{fig:VD_EC-R}, is non-tapered, except for the first (from photocathode, labelled PK) and last stage, and is optimised for maximum gain for single-photon detection. The total resistance of the voltage divider is $R = 3000~\mathrm{k\Omega}$.


The anode is grounded and a constant negative voltage, typically $V_\mathrm{1}= -900~\mathrm{V}$, is applied to the cathode resulting in an average gain of order $G=10^6$ electrons ($1~\mathrm{Me}$) and, in absence of a cathode photocurrent, in a power-supply current $i_\text{PS} = 300~\mathrm{\mu A}$ flowing through the resistors of the voltage divider.\footnote{Similar considerations to those made in this section can be done for tube HPK-R12699-M64, equipped with 10 dynodes and having a total resistance of $R = 2500~\mathrm{k\Omega}$.}

When there is incident light, the power supply current $i_\text{PS}$ is perturbed by the inter-electrode currents. This is particularly relevant at the last stage of multiplication, where the anode current $i_\text{A}$ can assume non-negligible values with respect to $i_\text{PS}$ and cause a voltage drop between the last dynode and the collection electrode. In order to keep a constant voltage $V_\mathrm{1}$ across the tube, the power supply current has to take a new value $i'_\text{PS} = i_\text{PS}+\Delta i_\text{PS}$, indicating with a prime the values in the presence of a photocurrent, to distinguish from those during operations with no signal. In addition, the voltage drop due to the inter-electrode currents has to be compensated by an increase of the inter-voltages at the first stages. Consequently, above a certain illumination rate, the proportionality between the anode current and the photocurrent, \ie the gain linearity, is lost. The relative variation of the gain takes the approximate relation $\Delta{G}/G \approx i_A/i_\text{PS}$ that is about 25\% for the highest illumination rate of the \rich detectors, corresponding to a photocurrent $i_\text{PK} \sim 80~\mathrm{pA}$ and $i_\text{A} = i_\text{PK}G \sim 80~\mathrm{\mu A}$ for $G=1~\mathrm{Me}$.

\begin{figure}[tb]
  \begin{center}
    \includegraphics[width=0.49\linewidth]{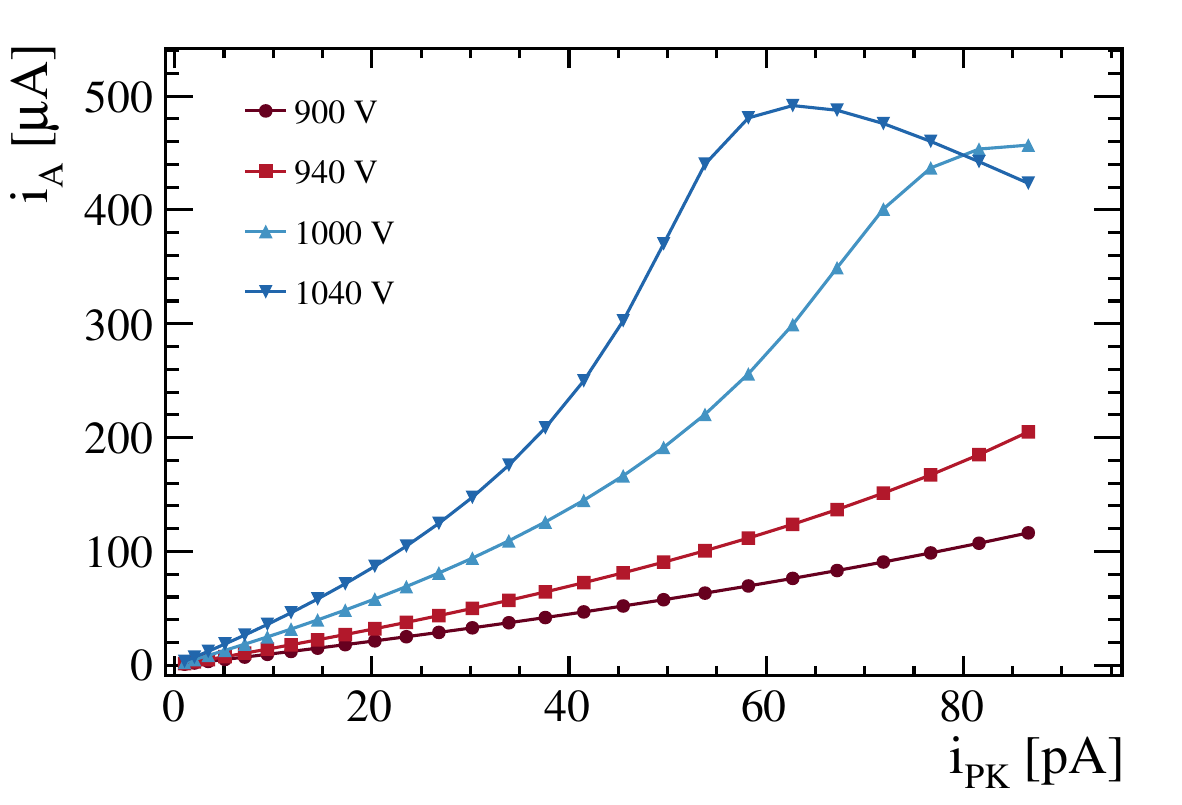}\put(-133,123){(a)}
     \includegraphics[width=0.49\linewidth]{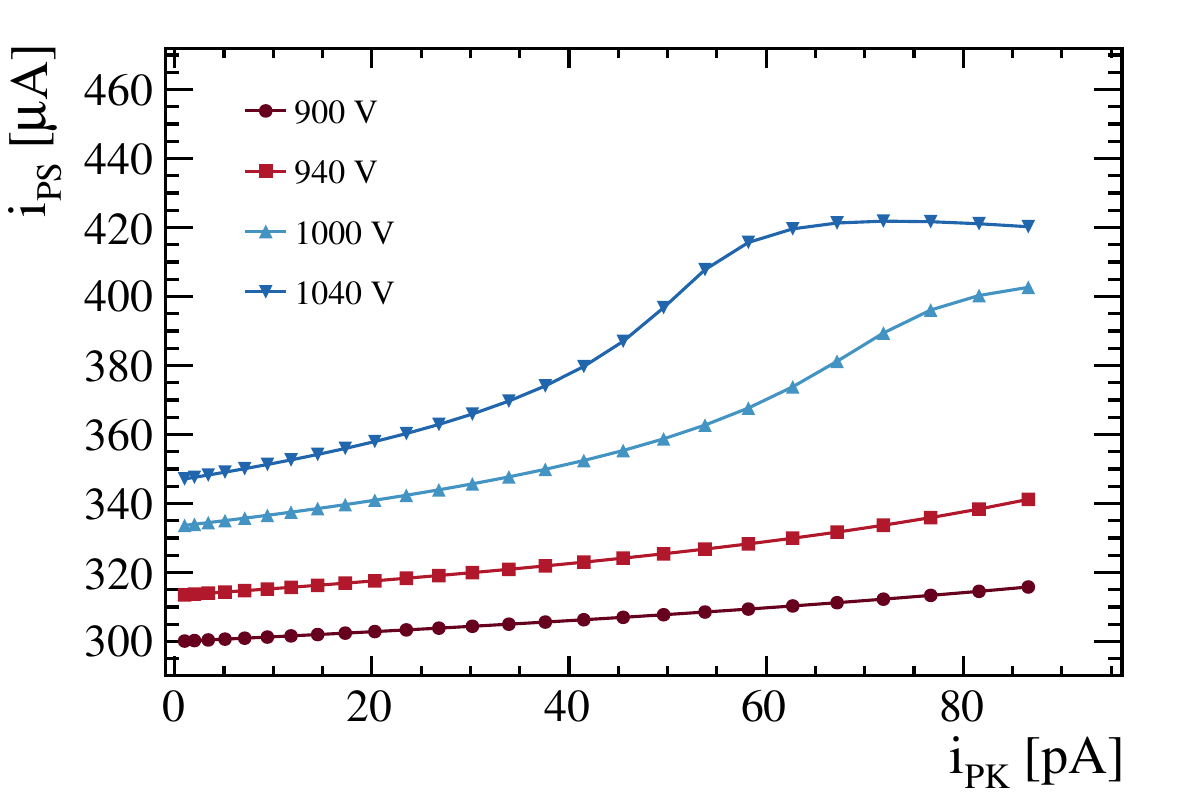}\put(-133,123){(b)}\\
     \includegraphics[width=0.49\linewidth]{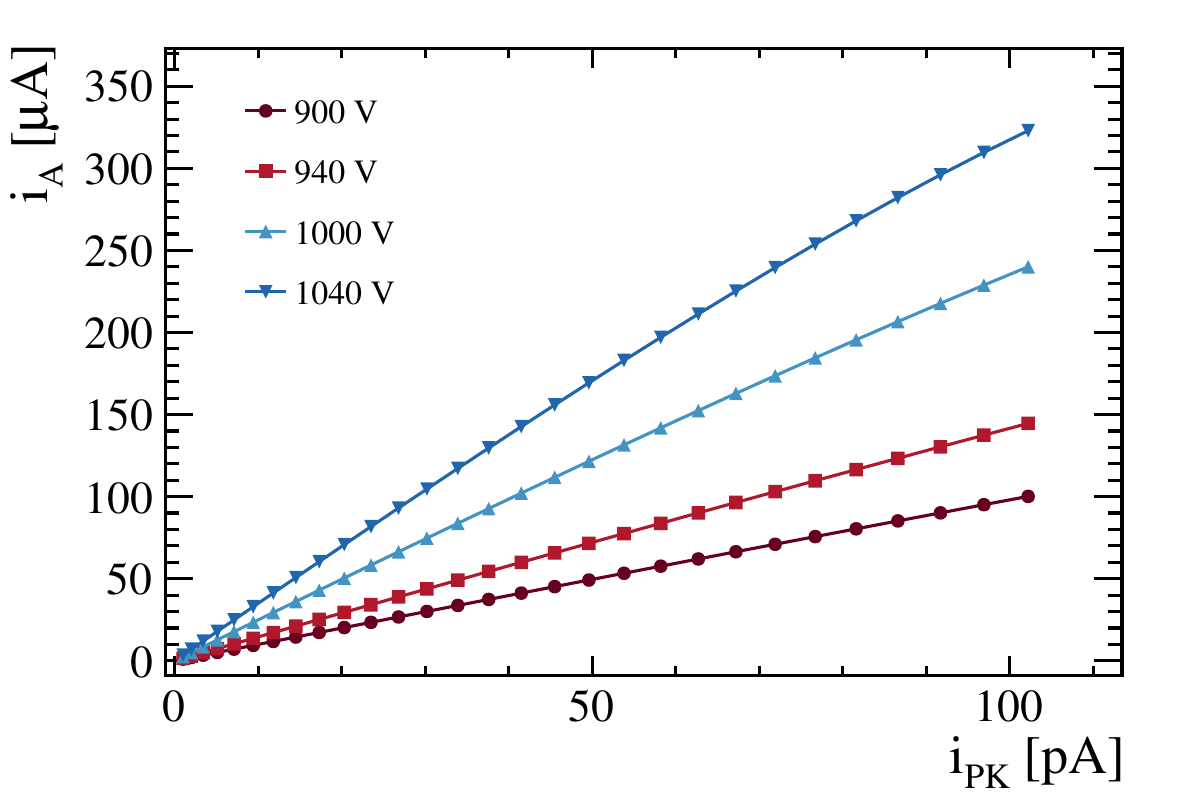}\put(-133,123){(c)}
     \includegraphics[width=0.49\linewidth]{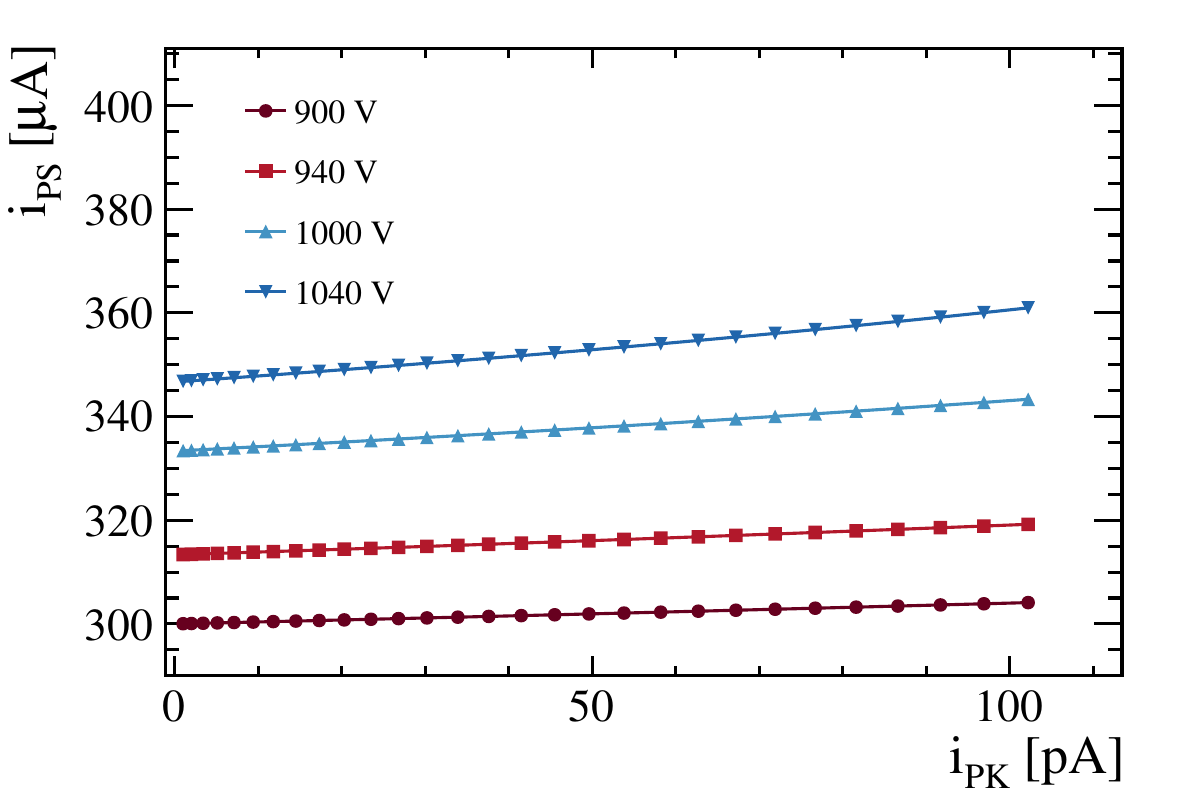}\put(-133,123){(d)}
    \vspace*{-0.5cm}
  \end{center}  
  \caption{Results of the LTspice simulation for different operating voltages of a $G=2.5~\mathrm{Me}$ HPK-R11265-M64 tube for (a) anode current $ i_A$ and (b) power-supply current $i_\text{PS}$ as a function of the photocurrent $i_\text{PK}$ for the standard powering configuration, showing the variations in the MaPMT gain. The so-called over-linearity is evident up to $i_\text{PK} \sim 60~\mathrm{pA}$, followed by a decrease of the output-signal amplitude due to the voltage drop between the last dynode and the anode degrading the collection efficiency. Figures (c) and (d) show the same results but with the last dynode powered independently, showing how the gain linearity is recovered especially for the curves below 1000 V that are closer to the working points used in the \rich detectors.}
  \label{fig:compareSPICE}
\end{figure}

To recover gain linearity the voltage divider of Fig.~\ref{fig:VD_EC-R} is equipped with auxiliary high-voltage power supplies for the last (Dy12) and last-but-one (Dy11) dynodes. In order to define the appropriate powering schemes for the \rich detectors, a purely electrical model of the voltage divider for tube HPK-R11265-M64 has been developed using LTspice~\cite{ltspice}. The model is used to simulate the ideal electrical behaviour of the biasing circuit. Experimentally the illumination rate is controlled through the average occupancy $O_\mathrm{av}$ at 40\mhz, where occupancy is here defined as the fraction of above-threshold pixels over the 64 channels of an MaPMT. Results are shown in Fig.~\ref{fig:compareSPICE} and Table \ref{table:1}, demonstrating that the independent powering of Dy12 preserves the linearity of the MaPMT response over a large range of occupancy values, reducing the maximum non linearity to $\Delta{G}/G \sim 2\%$.  

\begin{table}[tb]
\caption{Simulated voltage and current values as a function of occupancy $O_\mathrm{av}$ and corresponding photocurrent $i_\text{PK} $ with the photocathode and the last dynode Dy12 powered at the operational voltages of 1000 V and 100 V, respectively. $ i_\text{PS}$ ($i_\text{Dy12}$) is the current drawn by the main (auxiliary) power supply channel, and $V_\text{Dy11} \equiv V_\text{V2}$ is the voltage of the last-but-one dynode.}
\begin{center}
\begin{tabular}{ |c|c|c|c|c|c| } 
 \hline
 $O_\mathrm{av}$ [\%] & 0.01 & 1.00 & 10.00 & 20.00 & 30.00 \\ 
 \hline
 $ i_\text{PK}$ [pA] & 0.04 & 4.10 & 41.00 & 82.00 & 123.00 \\ 
  \hline
 $ i_\text{PS}$ [$\mu$A] & -333.34 & -333.49 & -334.91 & -336.57 & -338.31 \\ 
  \hline
 $ i_\text{Dy12}$ [$\mu$A] & -0.04 & -3.94 & -39.19 & -77.69 & -115.27 \\ 
  \hline
  $ V_\text{Dy11}$ [V] & 166.66 & 166.42 & 164.16 & 161.55 & 158.81 \\ 
  \hline
 $  i_A $ [$\mu$A] & 0.04 & 4.10 & 40.77 & 80.93 & 120.25 \\ 
  \hline
  \hline
 $G$ [$\times 10^6~\mathrm{Me}$] & 1.00 & 1.00 & 0.99 & 0.99 & 0.98 \\ 
 \hline
\end{tabular}
\end{center}
\label{table:1}
\end{table}

The powering of Dy12 keeps constant $V_\text{Dy12} \equiv V_\text{V3}$ and therefore compensates the voltage drop induced by the anode current by supplying a current $i_\text{Dy12} = (i'_\text{PS}-i_\text{PS})-i_A$. Hence the anode current can be determined measuring the power-supply currents as

\begin{equation}
\label{anodeCurrent}
    i_A = \Delta i_\text{PS} +  i_\text{Dy12},
\end{equation}

\noindent where $\Delta i_\text{PS}$ is the difference in the main power-supply current measured with and without a photocurrent. Table \ref{table:1} displays simulated voltage and current values as a function of occupancy and the corresponding photocurrent $i_\text{PK}$. The concept has been validated in laboratory tests using a pulsed laser light source. The linearity of the MaPMT response has been confirmed for occupancies up to 30\% when powering the last dynode.

Given the occupancy distribution of the \rich detectors, the powering of the last dynode is employed in \richone but not necessary in \richtwo. The operational working point is chosen to be at the lowest possible voltage allowing a three-sigma separation between the single-photon gain peak and the front-end electronics thresholds. This avoids exceeding the maximum anode-current limit $(i_A)_\mathrm{max} = 100~\mathrm{\mu A}$, defined by the tube specifications from Hamamatsu and monitored using Eq.~\ref{anodeCurrent}, and to prevent as much as possible MaPMT ageing. After an equalisation procedure based on high-voltage adjustments, the average MaPMT gain in operations is $G = 1.2~\mathrm{Me}$ while the front-end electronics threshold is set to $200~\mathrm{ke}$. Ageing of MaPMTs, over the long time-scale involved (fifteen years of operation in the \lhcb environment), is monitored through dedicated measurements of the gain variations that have been confirmed to be negligible for the dataset used in this paper. The practical implementation of the method and its use in the \rich system is detailed in the next section.

\section{Real-time luminosity monitors}
\label{sec:operations}

The average number of Cherenkov photons per event reaching the RICH photon detection planes is proportional to the number of above-threshold charged tracks passing through the gas radiators and thus to the number of primary $pp$ interactions. Since $i_\text{PK}$ is given by the rate of photoelectrons produced at the cathode, the photocurrent scales with the number of photons arriving at the photon detection arrays. Therefore, $i_A$ is also proportional to the number of interactions and can be expressed as

\begin{equation}
\label{eq:anodeCurrentECS}
 i_A = N_\gamma~\epsilon_\text{pde}~G~e~\nu_\text{int} + \delta,
\end{equation}

\noindent where $N_\gamma$ is the average number of Cherenkov photons per bunch-crossing reaching the MaPMT photocathode, $\epsilon_\text{pde}$ represents the average photon detection efficiency, $G$ denotes the single photon average gain, $e$ is the electric charge, $\nu_\text{int}$ is the interaction rate and $\delta$ represents the contribution to the anode current from other sources such as random internal instrumental noise or dark counts.

The interaction rate $\nu_\text{int}$, to  first order, is given by $\nu_\text{int} = N_{bb} \cdot f_r$, with $N_{bb}$ being the number of colliding bunches at the \lhcb interaction point and $f_r = 11.245\khz$ is the \lhc revolution frequency. The instantaneous luminosity can be written as 

\begin{equation}
\label{eq:lumiLHCb}
    \lum =  \nu_\text{int} \; \frac{\mu_{\text{LHCb}}}{\sigma_{\text{LHCb}}},  
\end{equation}

\noindent where $\mu_{\text{LHCb}}$ is the average number of visible collisions in \lhcb, defined as events in which at least one charged track produces hits through the full experiment acceptance. The value $\sigma_{\text{LHCb}}$ is the cross-section for these $pp$ collisions, whose value at 13\tev has been determined~\cite{LHCb-PAPER-2018-003}. The value $N_\gamma$ in Eq.~\ref{eq:anodeCurrentECS} is proportional to $\mu_{\text{LHCb}}$ and therefore the anode current serves as a proxy for the instantaneous luminosity, provided the gain is stable as a function of rate and occupancy. 

\begin{figure}[tb]
  \begin{center}
    \includegraphics[width=0.49\linewidth]{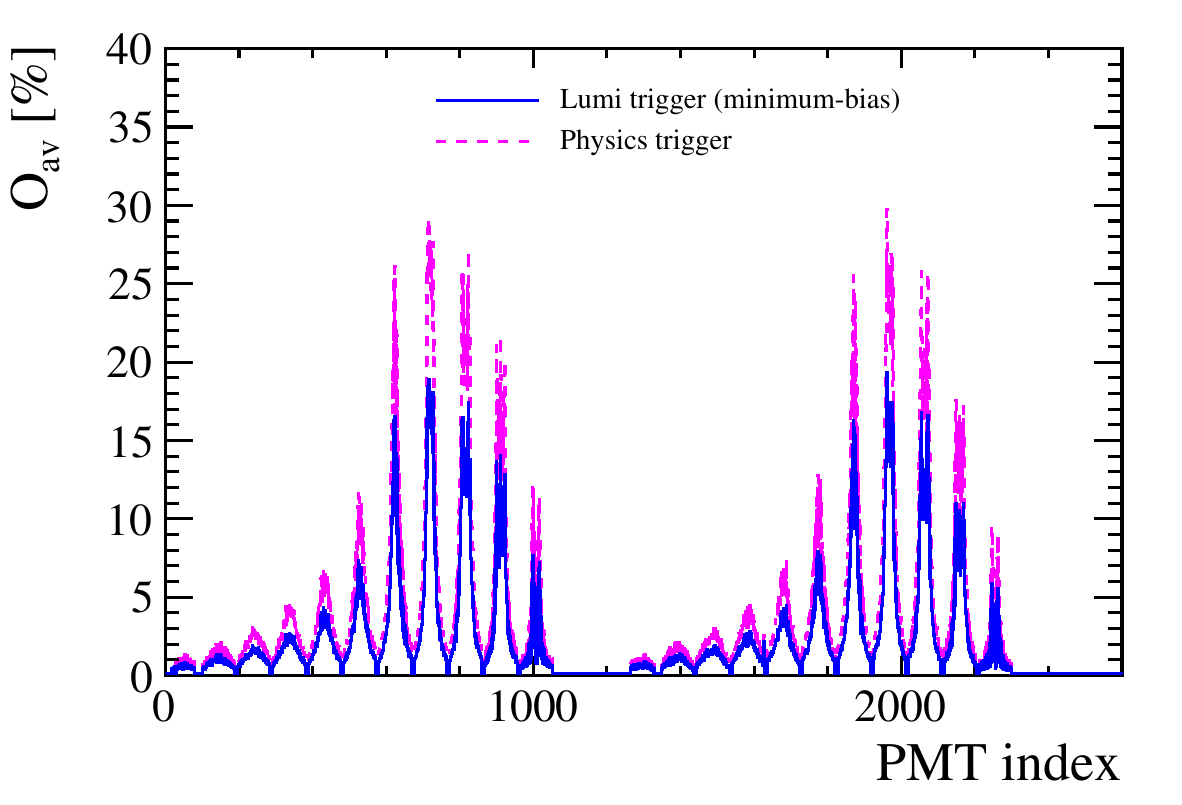}\put(-180,123){(a)}
    \includegraphics[width=0.49\linewidth]{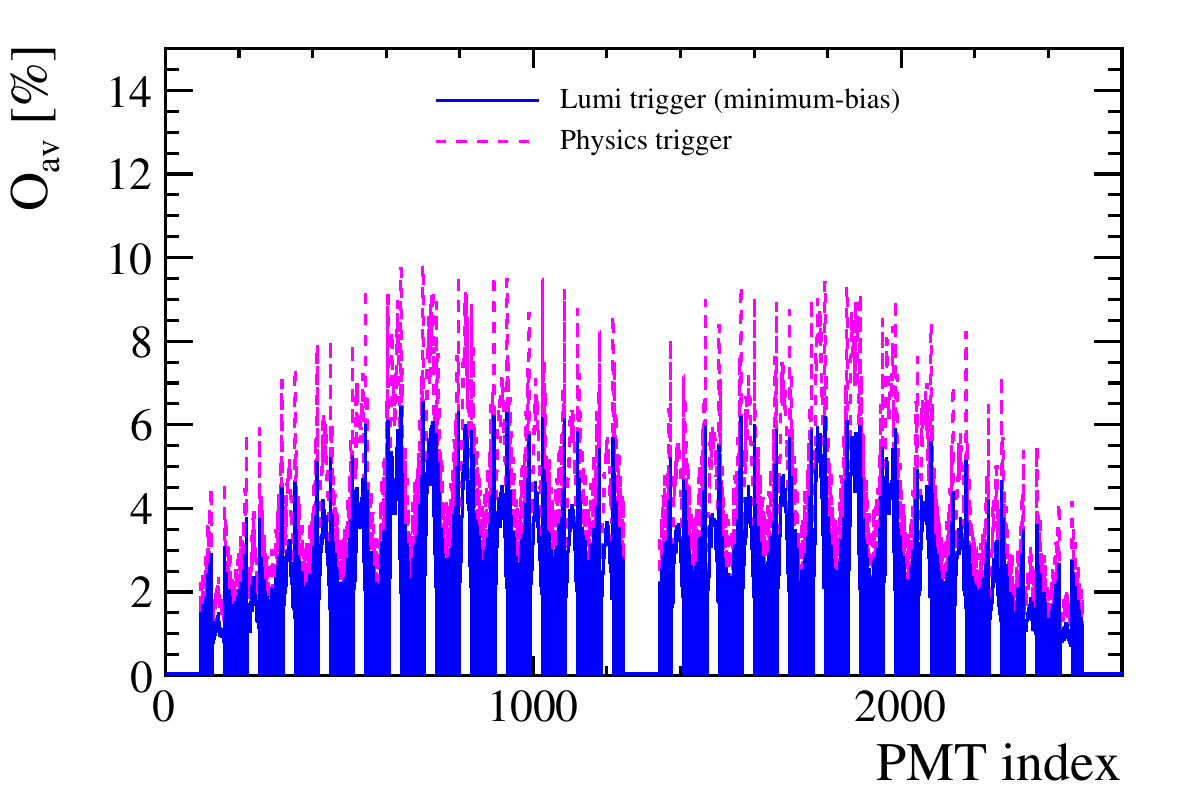}\put(-180,123){(b)} 
    \vspace*{-0.5cm}
  \end{center}
  \caption{Occupancy distributions for (a) \richone and (b) \richtwo as a function of MaPMT index number. The magenta dashed curves correspond to physics-biased events while the blue curves correspond to minimum-bias events.}
  \label{fig:sagradaFamilia}
\end{figure}

The occupancy is measured at nominal luminosity \lum = $2 \times 10^{33}~\mathrm{cm}^{-2}~\mathrm{s}^{-1}$ and is shown in Fig.~\ref{fig:sagradaFamilia}. The detected peak occupancy for physics-biased events, defined at first order as containing a $b$- or $c$-hadron, is approximately 30\% in \richone, with this number being relevant for the ability to distinguish charged hadrons for data analyses. Conversely, minimum-bias occupancy, determined by randomly selecting events corresponding to colliding bunches, results in an occupancy of only 20\% in \richone. The unbiased events are more relevant from the operational point of view, giving an indicator of the average activity per colliding bunch crossing on the photon detection planes. In \richtwo, the physics-biased and minimum-bias peak occupancies are of order 10\% and 6\%, respectively. 

Based on the considerations reported in Sec.~\ref{sec:spice}, the dynode bias mode is implemented for all the MaPMTs installed in the \richone detector. While maintaining a stable gain, the powering of Dy12 also allows the exploitation of Eq.~\ref{anodeCurrent}, namely 

\begin{equation}
\label{iLumiRichFormula}
    i_{\text{LumiRICH}} \equiv i_A = \Delta i_{\text{PS}} + i_{\text{Dy12}},
\end{equation}

\noindent as real-time instantaneous luminosity monitors. In \richtwo only PK is supplied, while the Dy12 channel is used in monitoring mode, \ie $i_{\text{Dy12}}=0$.

The power-supply currents are measured through the \rich HV system, employing common-floating-ground A1538DN CAEN boards. The monitoring period is of order 10 ms corresponding to about 100 \lhc orbits, hence the sensitivity is on average values rather than instantaneous currents. The current monitoring of the channels has a typical accuracy of $\pm 2\% \pm 3~ \mu\text{A}$ as reported by the manufacturer. Given a gain of order $1~\mathrm{Me}$ and a 20\% detected occupancy per event in the higher occupancy region of \richone, a measurable $i_{\text{LumiRICH}}$ can be achieved in the corresponding PDMs starting from interaction rates above 100\khz. In order to reach a required online precision of $\pm 5\%$ in all the 272 PDMs of the \rich detectors, an interaction rate above 1\mhz is required.\footnote{During 2023 operations, the number of colliding bunches in \lhcb with \lhc in full machine mode was 1821, thus corresponding to approximately 20\mhz.} This corresponds to an \lhc filling scheme comprising about 90 colliding bunches in \lhcb.
 
\begin{figure}[tb]
  \begin{center}
    \includegraphics[width=0.4\linewidth]{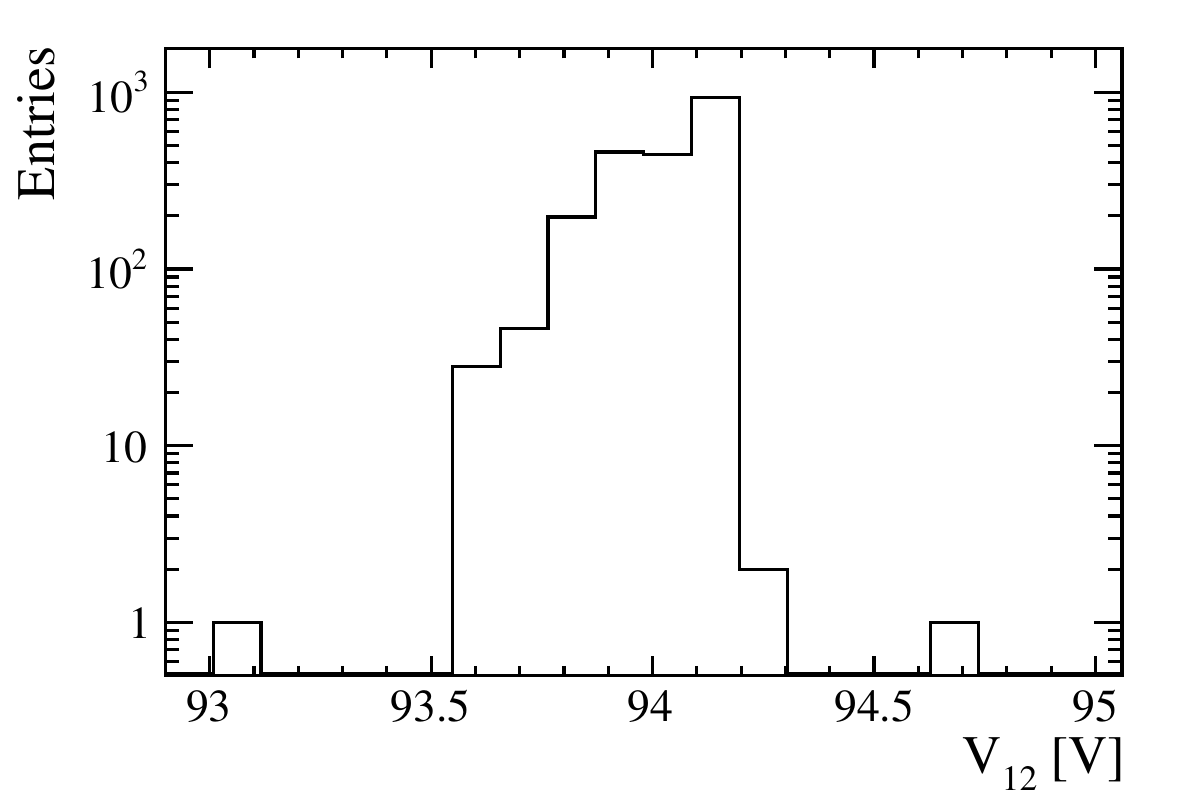}\put(-37,100){(a)}
    \includegraphics[width=0.5\linewidth]{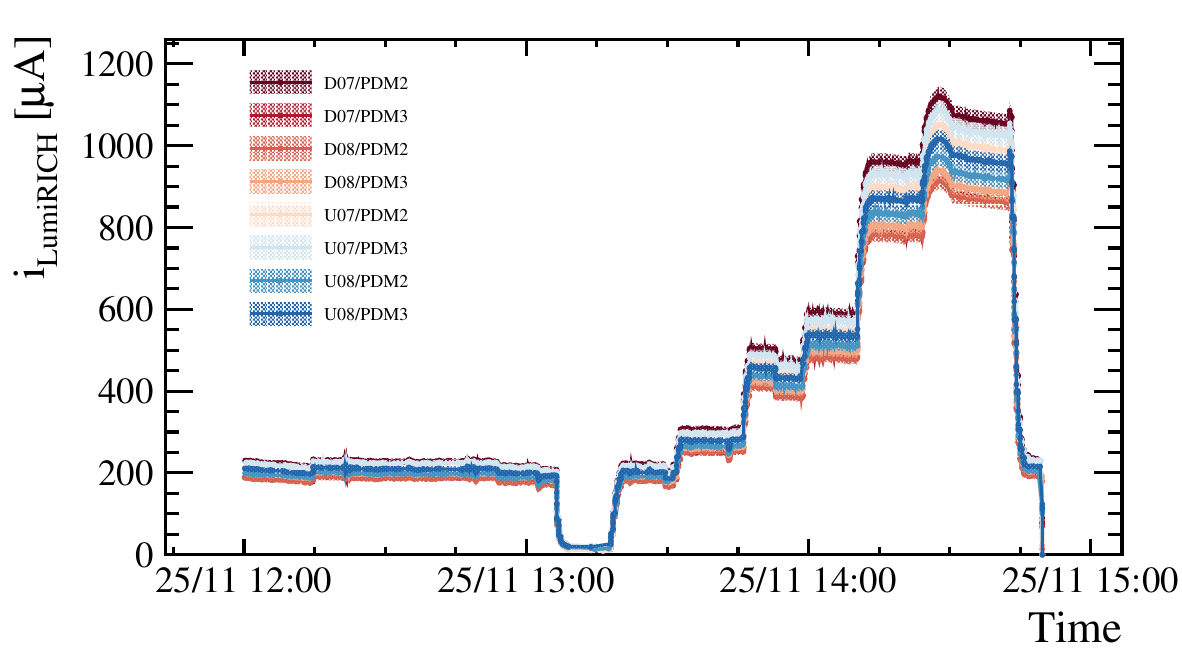}\put(-31,100){(b)}
    \vspace*{-0.5cm}
  \end{center}
  \caption{
    (a) The distribution of the last dynode voltage of a single PDM over the course of 2023 operations and (b) the trends of $i_{\text{LumiRICH}}$ for the eight PDMs in the high-occupancy region of \richone during a luminosity scan. Note that an excess around 14:30 is visible, indicating that beams were already colliding head-on. The variations in the absolute scale of these monitors is due to the different geometrical acceptances of the PDMs.}
  \label{fig:ecsOperations}
\end{figure}

The HV of the \rich detectors is steered by the Experiment Control System (\ecs), a set of \lhcb-specific components based on the JCOP framework and the WinCC-OA SCADA tool~\cite{wincc}. The operations of the HV system are fully automated and depend on the \lhc accelerator mode and the beam modes during an \lhc operational cycle~\cite{Brüning:691955}. The beam mode drives the actions of the HV system as follows: at the injection handshake between the \lhc and \lhcb, the high-voltage channels are turned off for safety to avoid damage to the photon detectors induced by possible showers of particles during the beams injection. When all trains are injected and the \lhc enters the beam-mode ramp to reach the target energy per beam of 6.8\tev, the photocathode channels are switched on, while the dynode channels are kept in monitoring mode. The \lhc takes about 30 minutes to accelerate the protons to the flat-top energy. After this so-called warm-up period for the MaPMTs, allowing stabilisation of the dark-count rate after switch-on, the dynode voltages are read out and the corresponding settings are saved in the \ecs Configuration Database. The advantage of reading in a measured set of dynode voltages, instead of applying the ones computed from the voltage divider ratio, lies in the possibility of factorising the tolerances of the voltage divider resistors and of incorporating first-order corrections to possible hysteresis of the power supplies due to environmental changes between one \lhc fill and the next. The spread in the voltage settings for Dy12 over the course of the 2023 run is shown in Fig.~\ref{fig:ecsOperations} (a) for a channel powering Dy12 of a single PDM. The standard deviation of the distribution is 0.132 V. If not corrected, an incorrect voltage setting would bias $i_{\text{Dy12}}$ of $\mathcal{O}(10~\mu\text{A})$. When the \lhc cycle reaches the so-called squeeze state, \ie the focusing of the beams at the interaction points starts but the beams are still separated, the \richone dynodes are powered. After a delay of two minutes, the baseline $i_{\text{PS}}$, $i_{\text{Dy12}}$ and $i_{\text{A}}$ currents are computed. This provides the reference values to compute the values entering Eq.~\ref{iLumiRichFormula} when stable beams, and therefore collisions, are declared. The term $\delta$ in Eq.~\ref{eq:anodeCurrentECS} due to dark counts is subtracted, and biases in the measured currents are corrected. 

Trending plots of $i_{\text{LumiRICH}}$ for the eight PDMs in the high-occupancy region of \richone are shown in Fig.~\ref{fig:ecsOperations} (b) to illustrate the sensitivity of this variable during a luminosity scan. The discontinuities are proportional to the steps in instantaneous luminosity. The visible number of interactions in \lhcb was increased from approximately 0.1 to 6. Data correspond to \lhc fill 8484, where 1735 bunches were colliding.

\begin{figure}[tb]
  \begin{center}
    \includegraphics[width=0.49\linewidth]{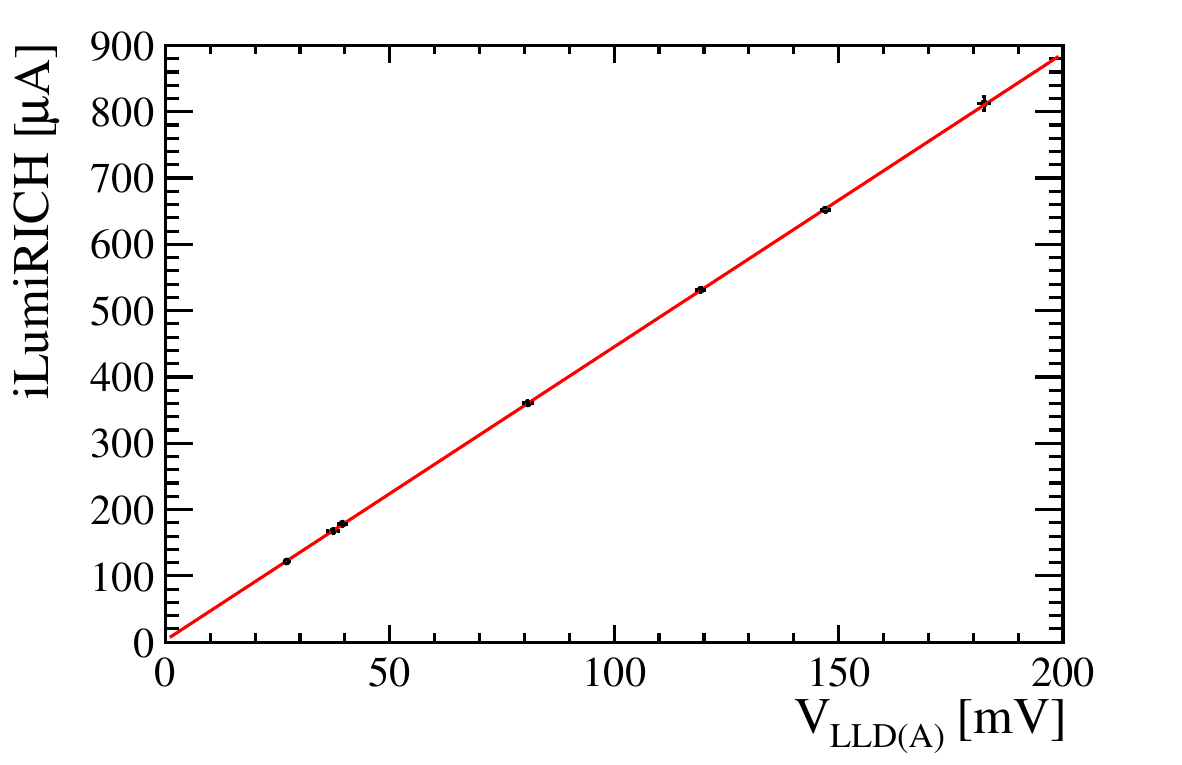}\put(-180,123){(a)}
    \includegraphics[width=0.49\linewidth]{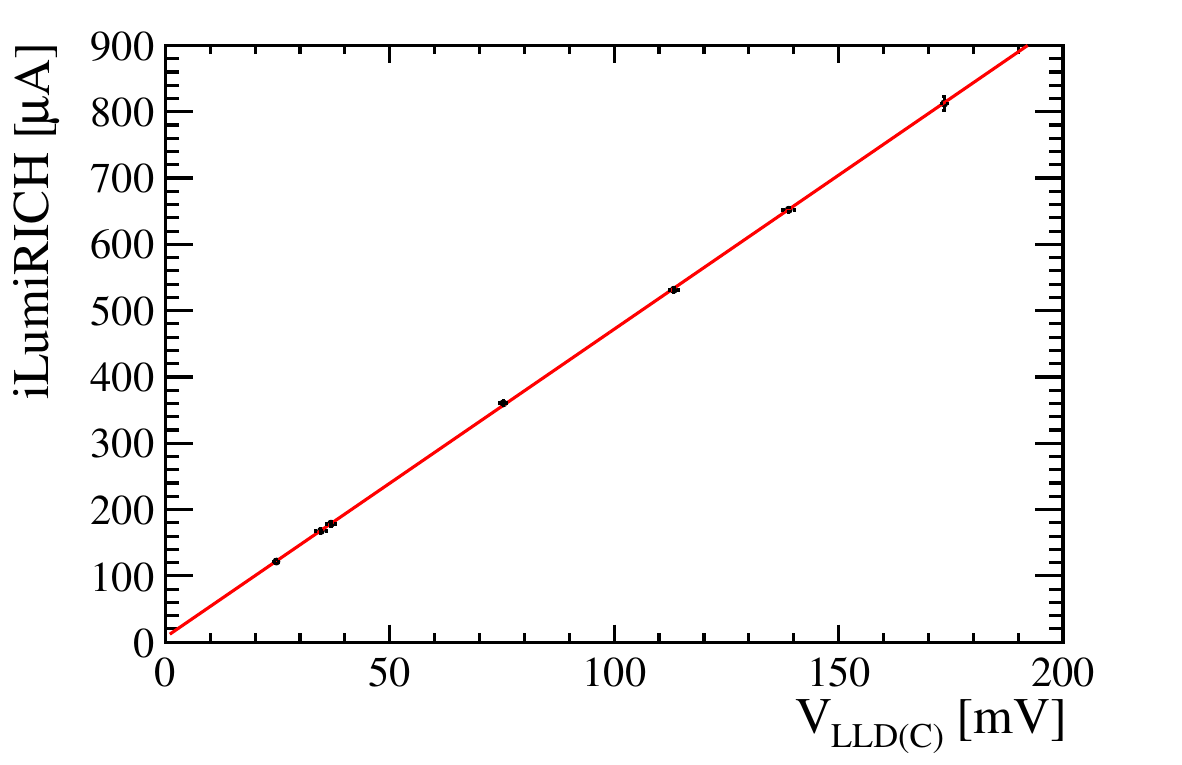}\put(-180,123){(b)}
    \vspace*{-0.5cm}
  \end{center}
  \caption{Correlation between MaPMT anode currents and the LLD voltages for \richtwo (a) A-side and (b) C-side. The baseline values are subtracted before the linear fit is performed. The fitted offset and slope values are $2.8 \pm 3.3~\mu\mathrm{A}$ and $4.43 \pm 0.04~\mu\mathrm{A}~\mathrm{mV}^{-1}$ for the A-side, and $7.5 \pm 3.5~\mu\mathrm{A}$ and $4.65 \pm 0.05~\mu\mathrm{A}~\mathrm{mV}^{-1}$ for the C-side.}
  \label{fig:LLDVsiLumiRich}
\end{figure}

Other quantities monitored through the \ecs that can be used for the luminosity measurement are the analogue outputs of the light leak detector (LLD), a safety system to check the tightness of the MaPMT enclosures to environmental light. Each LLD module is equipped with a photomultiplier tube whose output signals are discriminated, generating NIM pulses that are converted into TTL format and integrated over $\order{(100~\mathrm{\mus})}$ time intervals in a custom readout module. Two of the LLD modules, installed either side of \richtwo (the so-called A- and C-sides), are facing the radiator and have the greatest sensitivity. Similarly to the concept of Eq.~\ref{eq:anodeCurrentECS}, the LLD signal is proportional to the instantaneous luminosity. The calibrated output voltage ($V_\mathrm{LLD}$) can provide luminosity to the \lhc even when the MaPMTs are off, \eg during machine development periods. The correlation between MaPMT anode currents and the LLD output is shown in Fig.~\ref{fig:LLDVsiLumiRich}. 

One of the main advantages of the real-time monitors is their independence from the status of the \lhcb data-acquisition system. In order to find the proportionality constant between $i_{\text{LumiRICH}}$, $V_\text{LLD}$ and the instantaneous luminosity, the number of detected hits in the \rich photon detector planes can be employed, as reported in the following sections. 

\section{Hit counters and modelling of the opto-electronics chain}
\label{sec:sim}

The luminosity is measured using proxy variables, hereafter referred to as luminosity counters, for which the average number of visible collisions ($\mu_{\text{vis}}$) and the visible cross-section ($\sigma_{\text{vis}}$) are defined. For a given luminosity counter, defining $\varepsilon$ as the term factorising its geometric acceptance and detection efficiency, $\mu_{\text{vis}}=\varepsilon \mu_{\text {LHCb}}$ is a relative value that can be measured by counting the visible interactions in a detector, while $\sigma_{\text{vis}}=\varepsilon \sigma_{\text {LHCb}}$ is the visible cross-section. The most common methods used to obtain a relative luminosity measurement and estimate $\mu_{\text {vis}}$ are the linearity method and the log-zero method. The former consists of measuring the mean value of the counter in a certain time interval corresponding to a total number of events $N$. This is the simplest approach and requires the linearity of the counter response with luminosity in the full luminosity range, such that, defining $n_i$ as the value of the counter in event $i$, 

\begin{equation}
    \mu_{\text {vis}} = \dfrac{\sum_i n_i}{N}.
\end{equation}

In the log-zero method, the number of events with zero hits $N_{0}$, defined  according to a threshold that is process-dependent, is used to determine

\begin{equation}
\label{eq:logZero}
    \mu_{\text {vis}} = -\log\text{P}(0) = -\log\dfrac{N_{0}}{N}.
\end{equation}

\noindent The log-zero method, being based on Poisson statistics, begins to lose accuracy when the number of empty events of the counter decreases. The choice between these two methods depends on the properties of the luminosity counter under investigation. 

The reconstruction of the \rich detectors is performed at the second stage of the \lhcb High-Level Trigger (HLT)~\cite{LHCb-TDR-016}, allowing the use of the number of hits as luminosity counters. Following the RICH opto-electronics chain, photon signals detected by the MaPMTs are digitised and recorded as hits with a single-detector-pixel level of spatial granularity. In contrast to the real-time signals discussed in the previous section which can be used for online monitoring and averaging interactions over multiple \lhc orbits, the number of hits contains more comprehensive information. Specifically, the number of hits for each pixel is available for every bunch-crossing identifier (BXID), enabling the measurement of the luminosity per bunch crossing. Furthermore, this approach allows to separate the contributions from the four beam-crossing types, namely beam-beam ($bb$), beam-empty ($be$), empty-beam ($eb$), and empty-empty ($ee$). In order to provide a luminosity measurement that is not biased by physics selections performed at the HLT level, bunch-crossing are randomly selected during data-taking by the \lhcb readout supervisor~\cite{10115510}, responsible for distributing timing and fast control commands to all sub-detectors, at a rate of 22.5, 3.0, 3.0 and 1.5\khz for $bb$, $be$, $eb$ and $ee$ bunch-crossing types, respectively. The RICH counters have been implemented at the second HLT stage in a dedicated algorithm that stores all luminosity counters, along with the necessary event information, in a compact data format, for each event selected by the luminosity trigger. These events are routed through the HLT farm in pass-through mode and are also used to assess the status of the \rich detectors during data-taking, \eg to monitor the ageing of MaPMTs. 

The selection of the number of pixels that defines a luminosity counter variable is based on an optimisation of both the number and stability of these observables. The \rich hits are firstly investigated using different \lhcb minimum bias simulation samples generated with various values of the average number of $pp$ interactions per bunch crossing. The full simulation of the \rich detectors is employed, including the complete geometry of the detectors and the MaPMT pixel properties such as gain and dark counts. A realistic electronics modelling is also implemented, allowing access to the number of hits after the digitisation of the signal. The beam conditions, including the average position and size of the interaction region, the crossing angle and the emittance, are incorporated at the generator level. Simulation samples with different interaction positions along the $x$, $y$, and $z$ axes and with both magnet polarity configurations are used to evaluate the robustness of the counters as a function of the beam-spot position. 

The simulation samples are used to evaluate the scaling of the occupancy of the \rich detectors as a function of the number of primary $pp$ collisions per event. As discussed in Sec.~\ref{sec:operations}, the peak occupancy at the average nominal pile-up of 5.5 primary vertices per colliding bunch crossing is approximately 20\%. This results in a non-negligible non-linearity in the high occupancy region of \richone, arising from events in which two or more photons hit the same MaPMT pixel. These multi-hits are indistinguishable from single-photon hits due to the binary readout of the \rich opto-electronics chain. At this peak occupancy, the probability to record at least two hits in the same pixel is approximately 2\%.

\begin{figure}[tb]
  \begin{center}
    \includegraphics[width=0.32\linewidth]{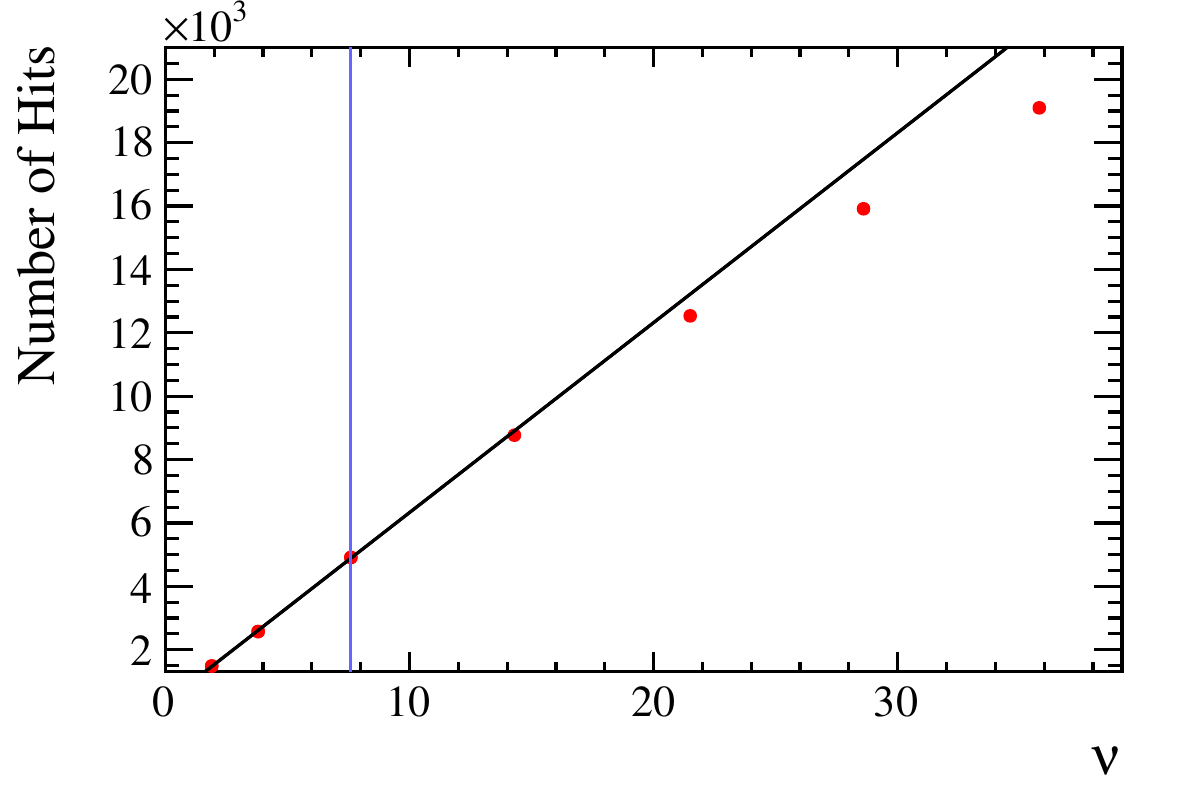}\put(-120,80){(a)}
     \includegraphics[width=0.32\linewidth]{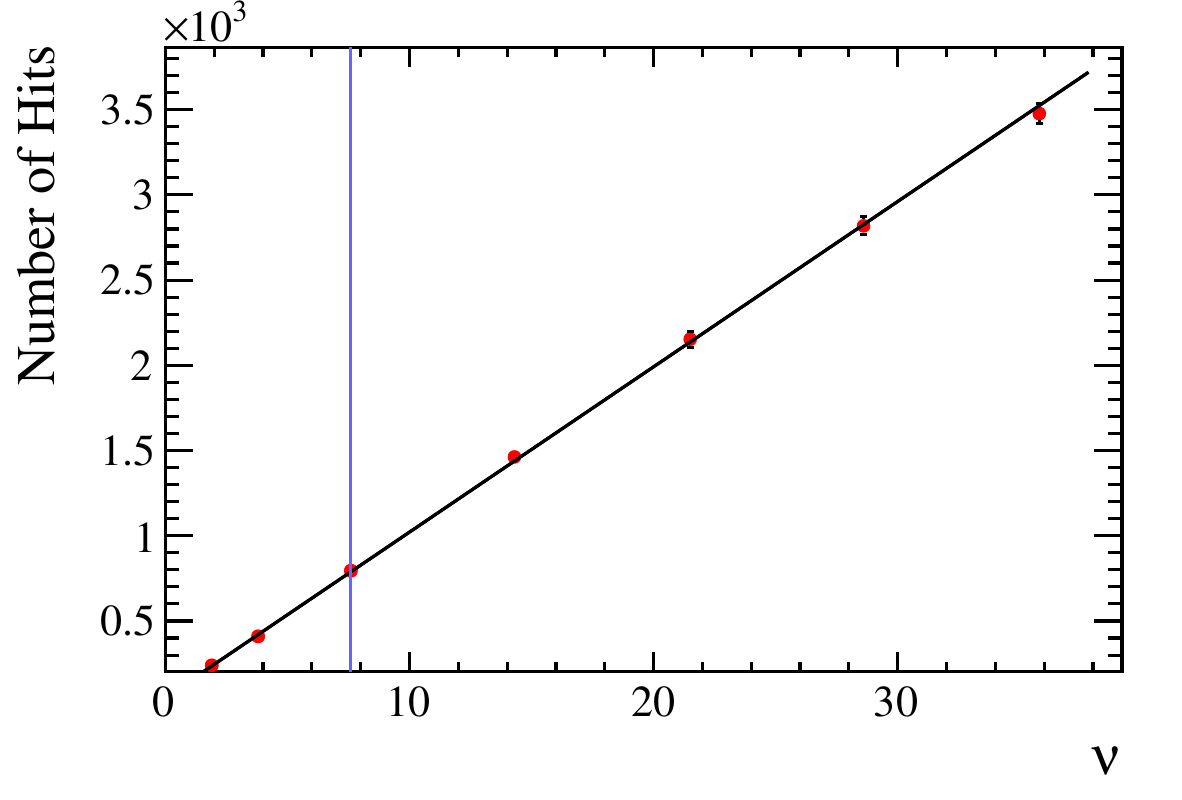}\put(-120,80){(b)}
     \includegraphics[width=0.32\linewidth]{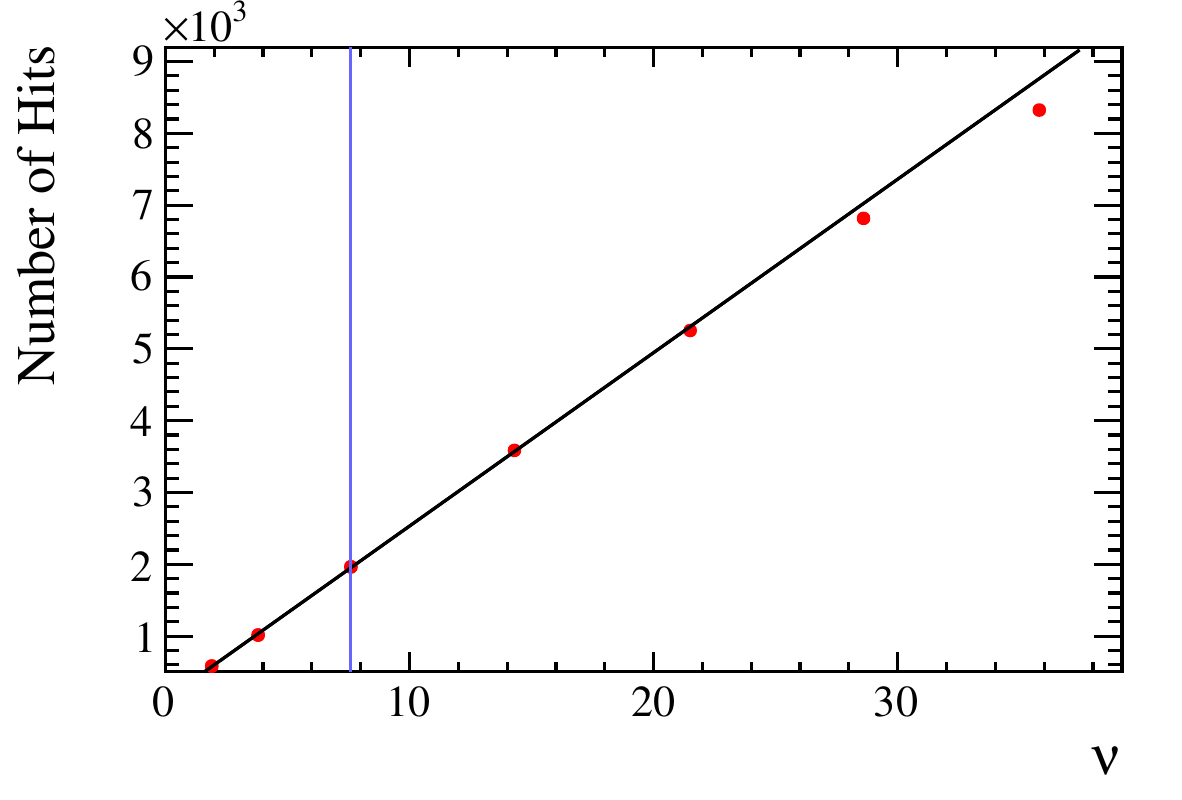}\put(-120,80){(c)}\\
    \includegraphics[width=0.32\linewidth]{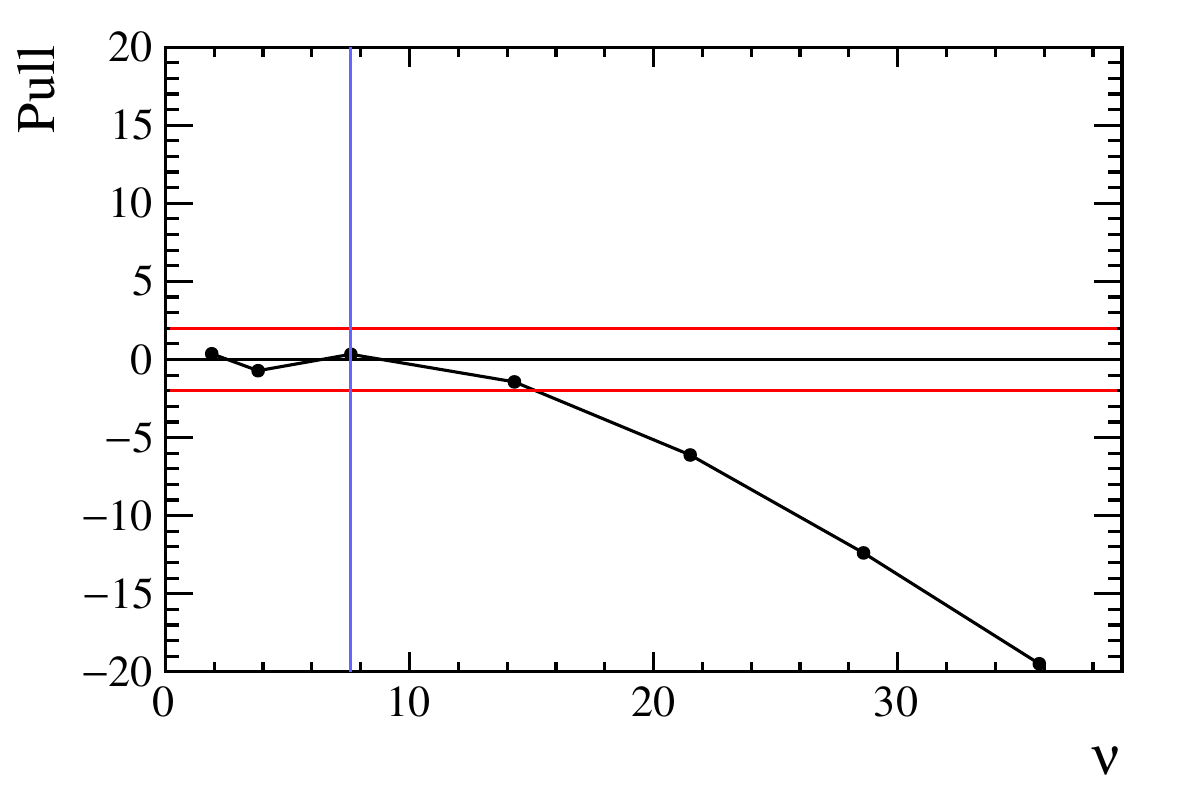}\put(-120,80){(d)}
    \includegraphics[width=0.32\linewidth]{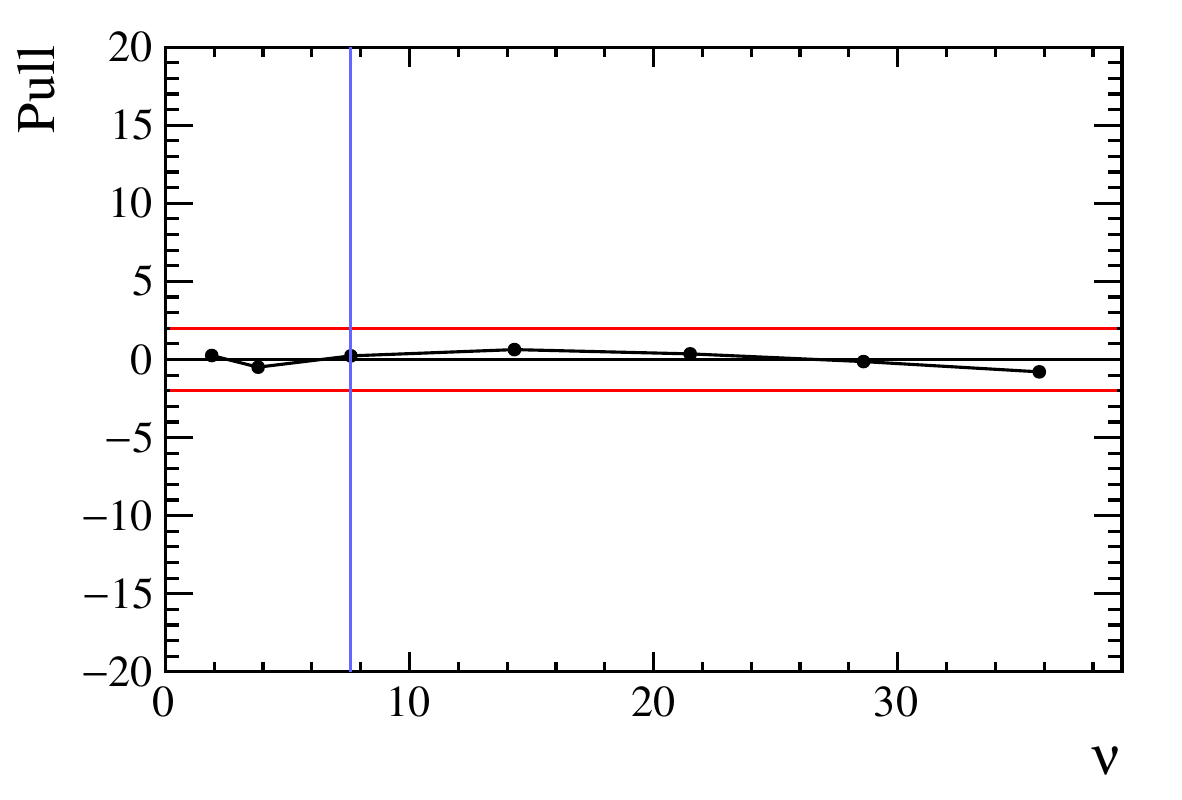}\put(-120,80){(e)}
    \includegraphics[width=0.32\linewidth]{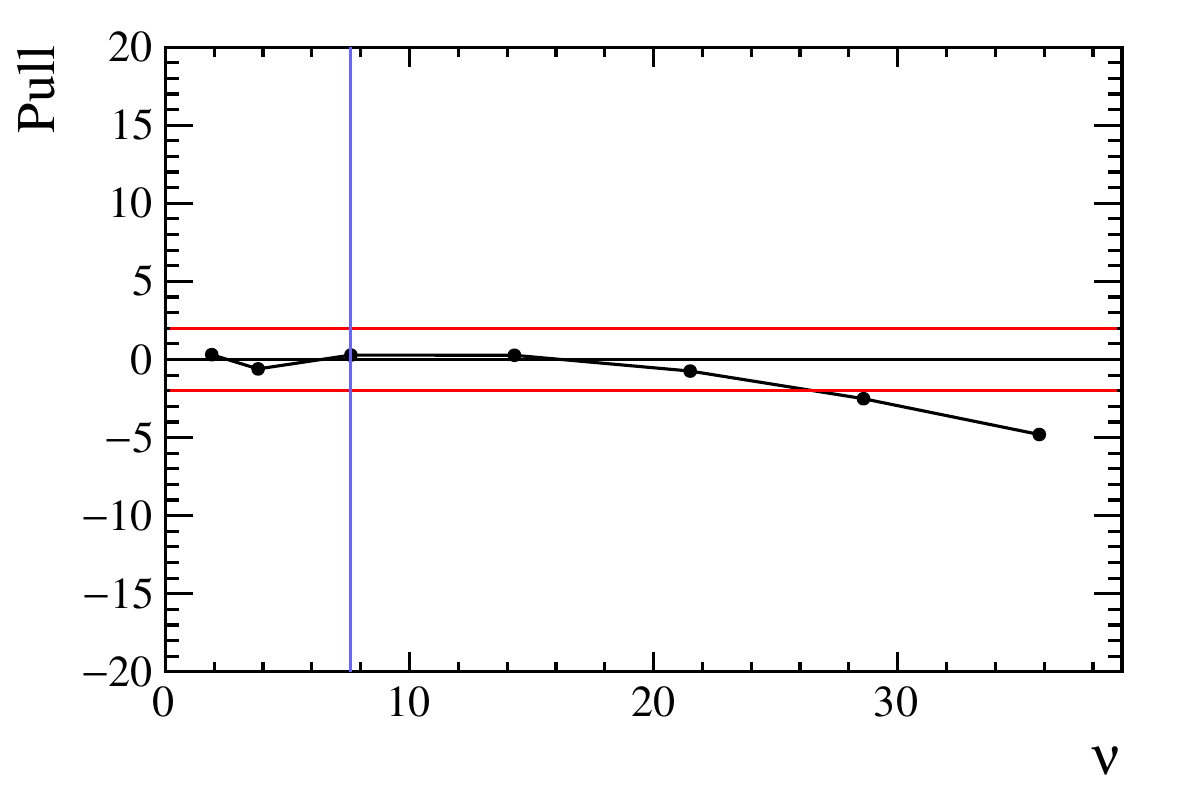}\put(-120,80){(f)}
    \vspace*{-0.5cm}
  \end{center}

  \caption{Mean number of hits as a function of the average number of primary $pp$ interactions $\nu$ for (a) \richone, (b) the outer regions of \richtwo, populated with two-inch HPK-R12699-M64s due to the lower occupancy, and (c) for the central regions of \richtwo. In the top row, the linear fit is performed only considering the first 3 points, the intercept is left free, as for $\nu = 0$, some hits are expected due to background noise. The pulls of the fits are shown in the bottom row, showing a significant deviation from linearity for $\nu > 10$ in \richone and for $\nu > 20$ in the inner region of \richtwo. Note that the average Run 3 pile-up corresponds to $\nu=7.6$, highlighted by the blue vertical line in the plots.}
  \label{fig:richregions}
\end{figure}

The distribution of the number of recorded hits is studied in different regions of the photon detector planes, as shown in Fig.~\ref{fig:richregions}. The mean of these distributions is plotted against the average number of $pp$ interactions per bunch crossing $\nu$, used in Monte-Carlo generators. The visible cross-section in \lhcb has been estimated from simulation, and the ratio between $\sigma_{\text{LHCb}}$ and the total $pp$ cross section is used to express the average number of visible collisions in \lhcb, $\mu_{\text{LHCb}}$, as a function of $\nu$ with the relation $\mu_{\text{LHCb}}=0.699\;\nu$.

\noindent As anticipated, the central region of \richone, with its higher occupancy, displays a deviation from linearity. In contrast, \richtwo generally exhibits lower occupancy and a hit count that is approximately linear with respect to $\nu$ up to values that are more than four times greater than nominal Run 3 conditions. These observations motivate the use of the log-zero method when determining the luminosity. This choice is adopted for all counters for consistency and to allow removal of biases arising from the most occupied events.

These exploratory studies have been performed before the start of Run 3 data taking and therefore employ a not-yet-tuned simulation. However, they incorporate effects seen during the 2022 and 2023 data-taking years, as described in the following sections. Following these studies, a \rich hit luminosity counter has been defined by the number of hits in a PDM. This choice results in a number of counters equal to the total number of PDMs in the RICH system, specifically, 128 hit counters for \richone (as the two outer columns have only four PDMs mounted) and 144 for \richtwo. Another advantage of this choice is the uniformity in the detector response, since each PDM encloses MaPMTs grouped by similar gains and powered by a common HV channel, providing a direct link with the corresponding real-time monitors described in Sec.~\ref{sec:operations}. Among the 272 available counters, a subset of 48 has been integrated into the offline luminosity data flow and is continuously stored to measure offline luminosity. The selection of stored counters is based on the PDMs providing the most precise and accurate luminosity estimates, while also optimising offline resources usage.

\section{Calibration of hit counters and luminosity determination}
\label{sec:VdM}

\begin{figure}[tb]
  \begin{center}
    \includegraphics[width=0.49\linewidth]{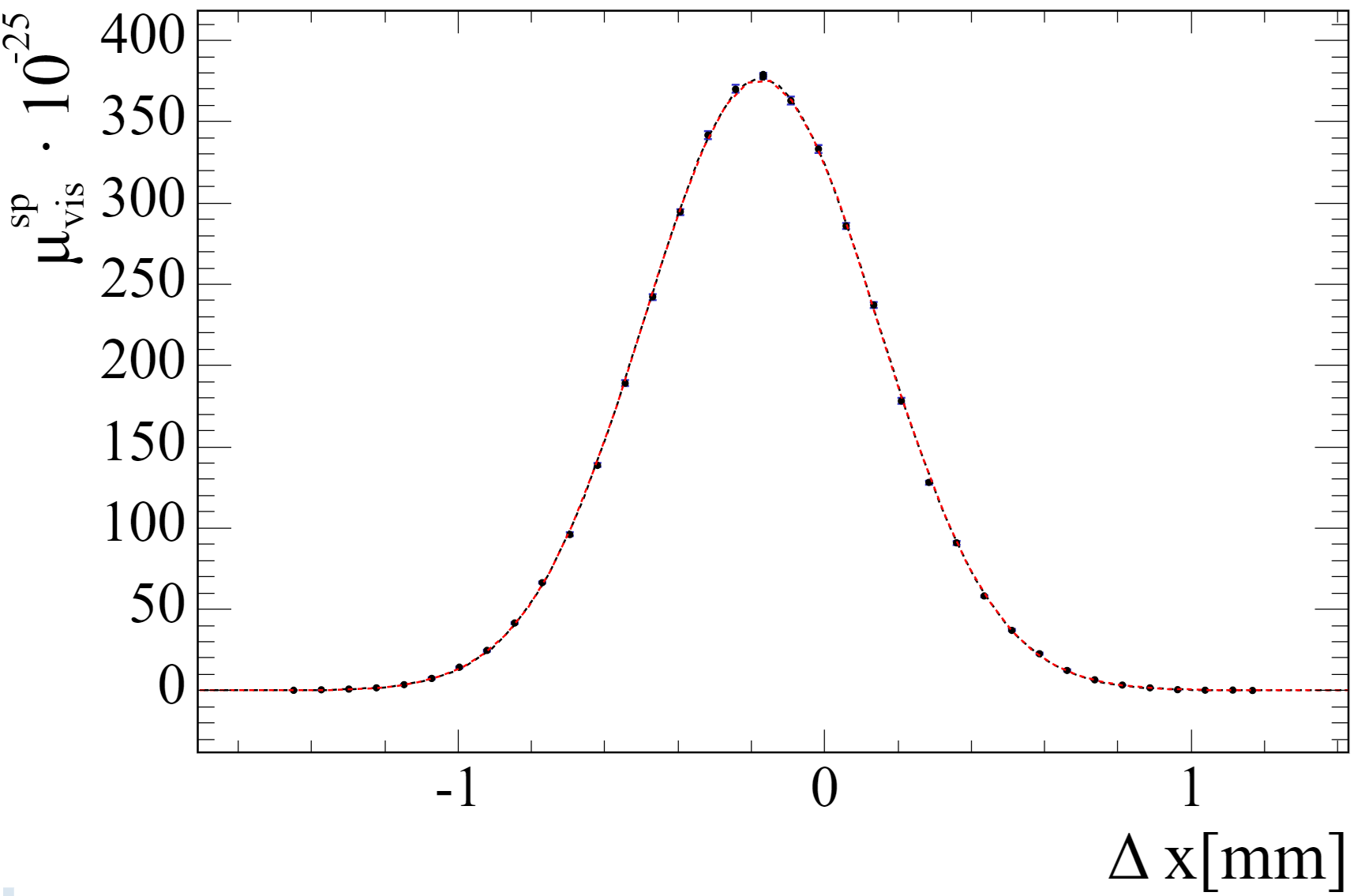}\put(-180,123){(a)}
    \includegraphics[width=0.49\linewidth]{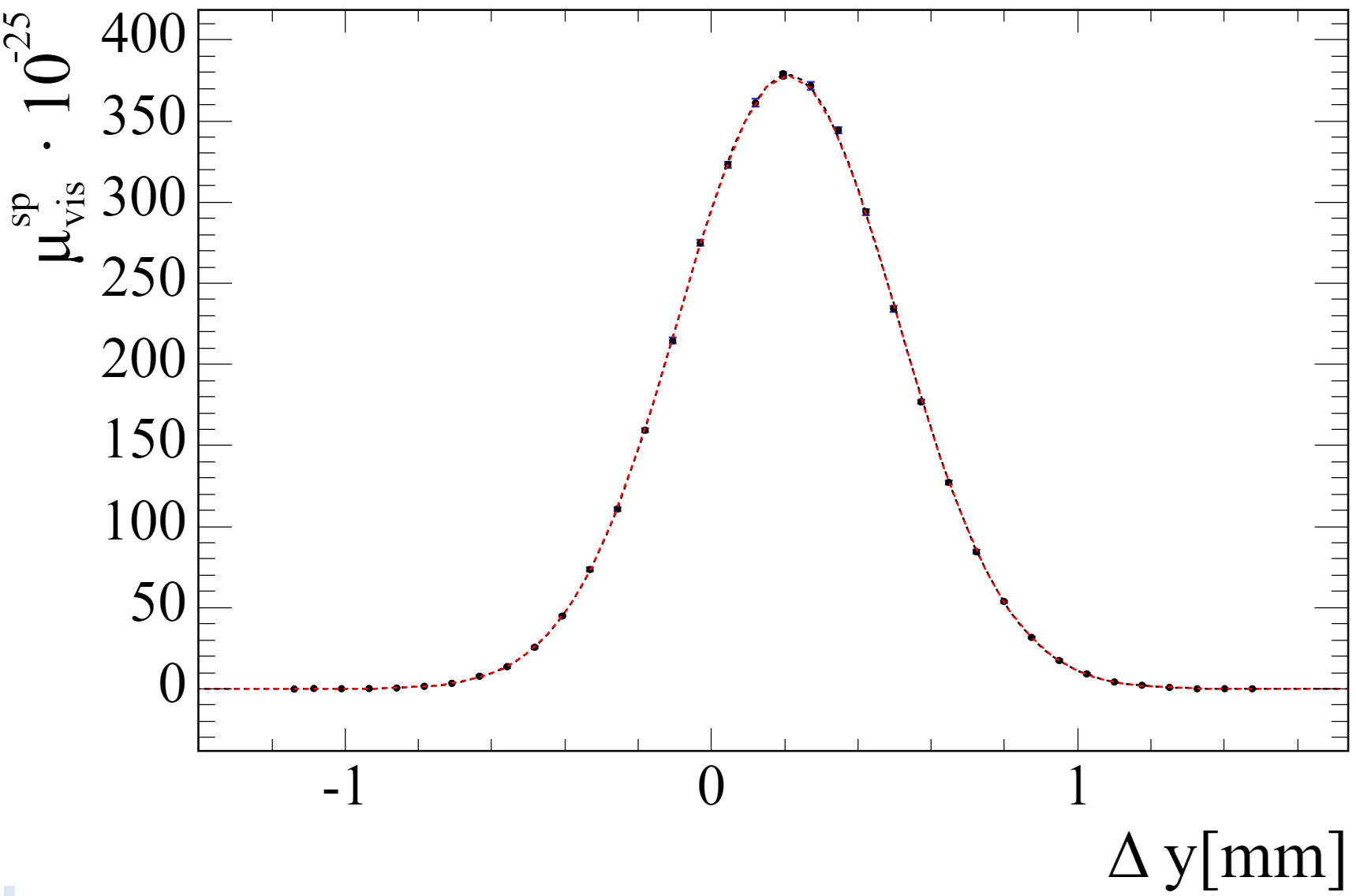}\put(-180,123){(b)}
    \vspace*{-0.5cm}
  \end{center}
  \caption{Profile of the $\mu_{\mathrm{vis}}^{\mathrm{sp}}$ parameter obtained by scanning along the (a) $x$ and (b) $y$ axes. The values are plotted against the relative displacement of the two colliding beams for a pair of colliding bunches. The shape is fitted with a Gaussian distribution. This fit is performed in the $x$ and $y$ transverse coordinates separately, assuming factorisation.}
  \label{fig:xyscan_fit}
\end{figure}

The van der Meer (vdM) scan technique~\cite{vanderMeer:1968zz,rubbia} is a direct method employed by all experiments at the \lhc to derive an absolute luminosity measurement, a crucial foundation for evaluating the instantaneous luminosity through counters calibrated with their respective cross-sections. Data are collected using the \rich system during vdM scans at \lhcb, concurrently with other subdetectors, with the aim of measuring the visible cross-section for each counter. A vdM scan can be performed in one or two dimensions by changing the beam separation at each step, the one-dimensional scan assuming factorisation of the beam profiles along the $x$ and $y$ axes, given that the transverse betatron oscillations are expected to be well decoupled in $x$ and $y$. Two-dimensional scans, introduced more recently, allow determination of the systematic uncertainty associated with this assumption by scanning the whole $x$-$y$ plane. The procedure to obtain the visible cross-section consists of evaluating the number of visible interactions $\mu_{\mathrm{vis}}$ in each vdM step which corresponds to a different overlap of the two colliding beams. The so-called specific average number of interactions ($\mu_{\mathrm{vis}}^{\mathrm{sp}}$) is the background-subtracted value of $\mu_{\mathrm{vis}}$ normalised by the quantity $N_{1}\cdot N_{2}$. For a colliding BXID with populations $N_1$ and $N_2$, the non-colliding background contribution due to the passage of single beams is estimated in every be, eb, and ee BXID by calculating $\mu$ using the log-zero method, and the average value for each BXID type is subtracted. The specific $\mu_{\mathrm{vis}}^{\mathrm{sp}}$ for step $i$ is defined as

\begin{equation}
\label{eq:muvis}
\mu_{\mathrm{vis}}^{\mathrm{sp}}\left(\Delta x_i, \Delta y_i\right)=\frac{\mu_{\mathrm{vis}}\left(\Delta x_i, \Delta y_i\right)- N_1 \textlangle \mu_{\mathrm{be}}/N_{1,\mathrm{be}} \textrangle - N_2 \textlangle \mu_{\mathrm{eb}}/N_{2,\mathrm{eb}}\textrangle + \textlangle \mu_{\mathrm{ee}} \textrangle }{N_1 \cdot N_2},
\end{equation}

\noindent where $\Delta x_i$ ($\Delta y_i$) is the displacement of the beams along $x$ ($y$) in step $i$.
The quantity $\mu_{\mathrm{vis}}$ is evaluated for each step of the vdM scan and for all hit counters using the log-zero method as for Eq.~\ref{eq:logZero}. The number of protons in each beam varies during the scan and is therefore evaluated for each step independently. The information from Direct-Current Current-Transformers~\cite{Chrin:2009zz} is used to measure the total beam current circulating in each \lhc ring and, simultaneously, the Fast Bunch Current Transformers~\cite{Dillingham:2010zz}, one per ring, provides a relative measure of the charges in each 25 ns \lhc bunch slot. The $x$ and $y$ profile obtained with a one-dimensional scan is shown in Fig.~\ref{fig:xyscan_fit} for a single hit counter. The shape is fitted using a Gaussian distribution for the signal and a constant for the background, the latter assumed uniform throughout the scan since the $ee$, $be$ and $eb$ contributions do not depend on the relative displacement of the beams. 

The value of the visible cross-section $\sigma_{\mathrm{vis}}$ for the case of two-dimensional scans, used for the results reported in this paper, can be obtained by using

\begin{equation}
\sigma_{\mathrm{vis}} =\int \mu_{\mathrm{vis}}^{\mathrm{sp}}(\Delta x, \Delta y) d \Delta x d \Delta y.
\end{equation}

\begin{figure}[tb]
  \begin{center}
    \includegraphics[width=0.49\linewidth]{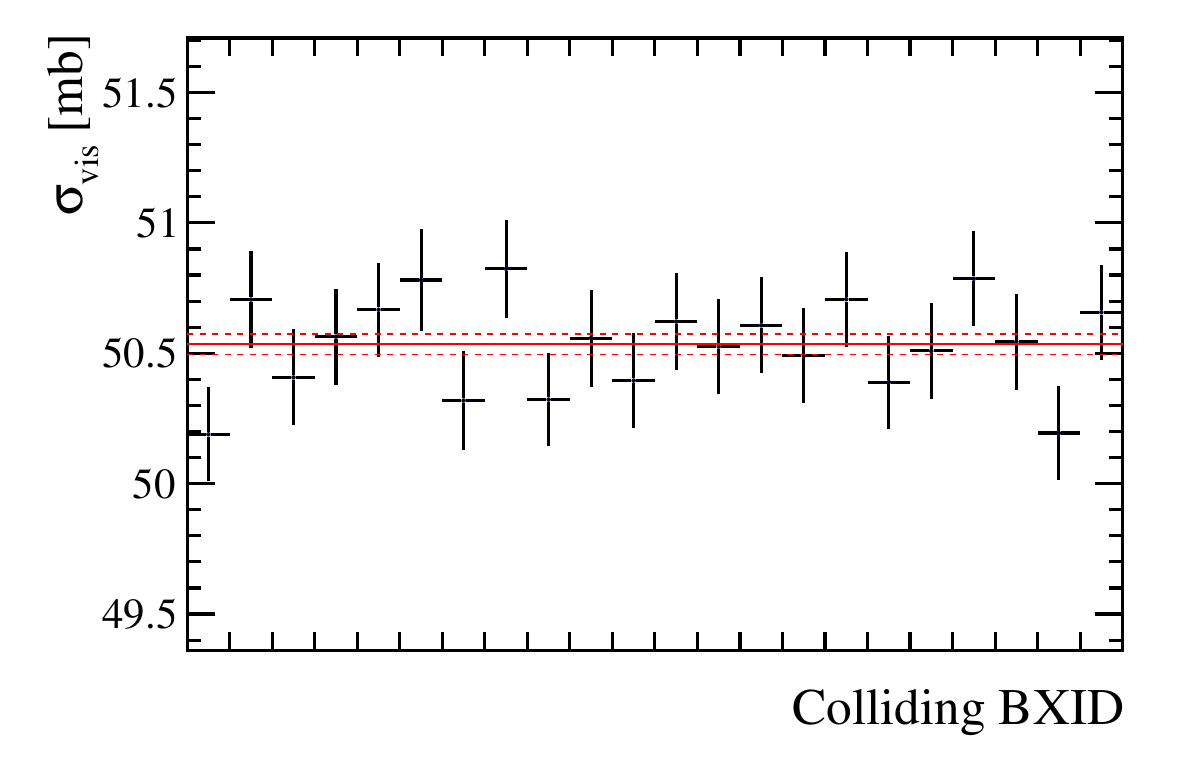}\put(-175,123){(a)}
    \includegraphics[width=0.49\linewidth]{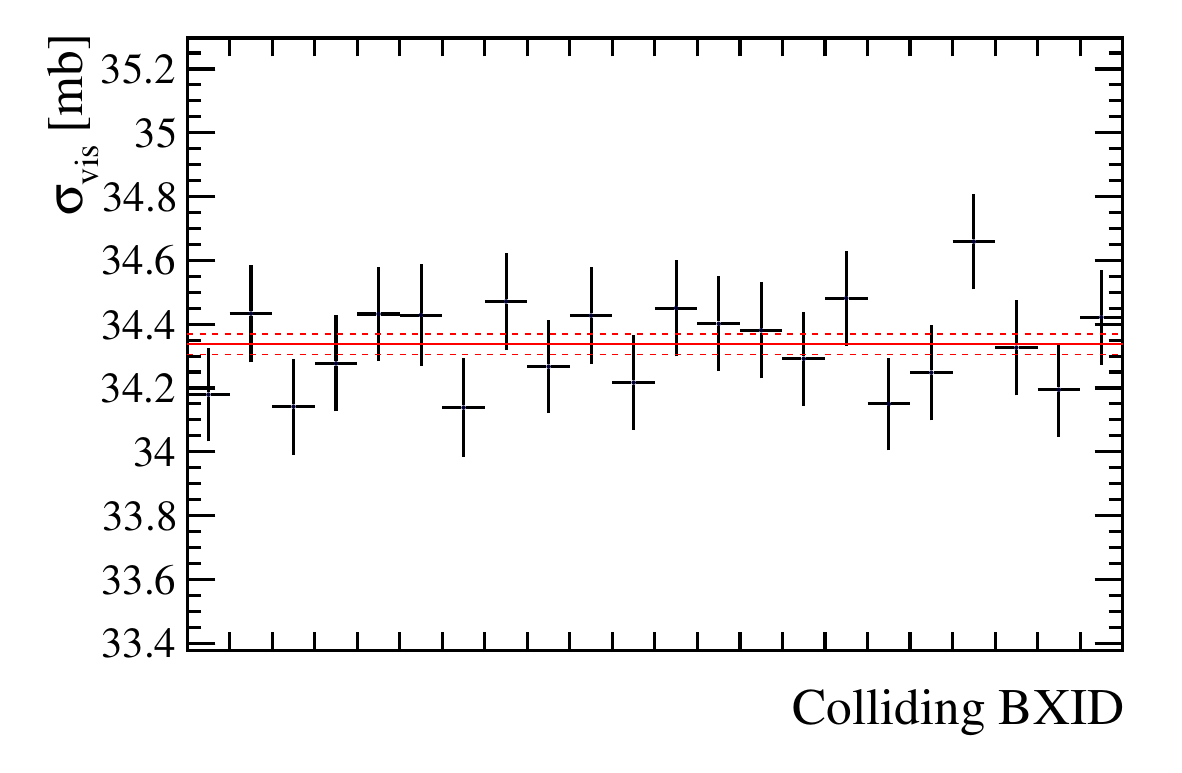}\put(-175,123){(b)}
    \vspace*{-0.5cm}
  \end{center}
  \caption{Visible cross-section values for each colliding BXID during the vdM scan for (a) one \richone hit counter in a high occupancy region and (b) one \richtwo hit counter in a high occupancy region, corresponding to average values across all BXIDs of $\sigma_{\mathrm{vis}} = 50.50  \pm 0.04 \mbarn$ and $\sigma_{\mathrm{vis}} = 34.34 \pm  0.03 \mbarn$, respectively, where the uncertainty is statistical only. The interval corresponding to one standard deviation is highlighted with red dashed lines.}
  \label{fig:meansigma}
\end{figure}

\noindent The cross-section is obtained for each colliding BXID as shown in Fig.~\ref{fig:meansigma}. 

\begin{figure}
    \centering
    \includegraphics[width=0.9\linewidth]{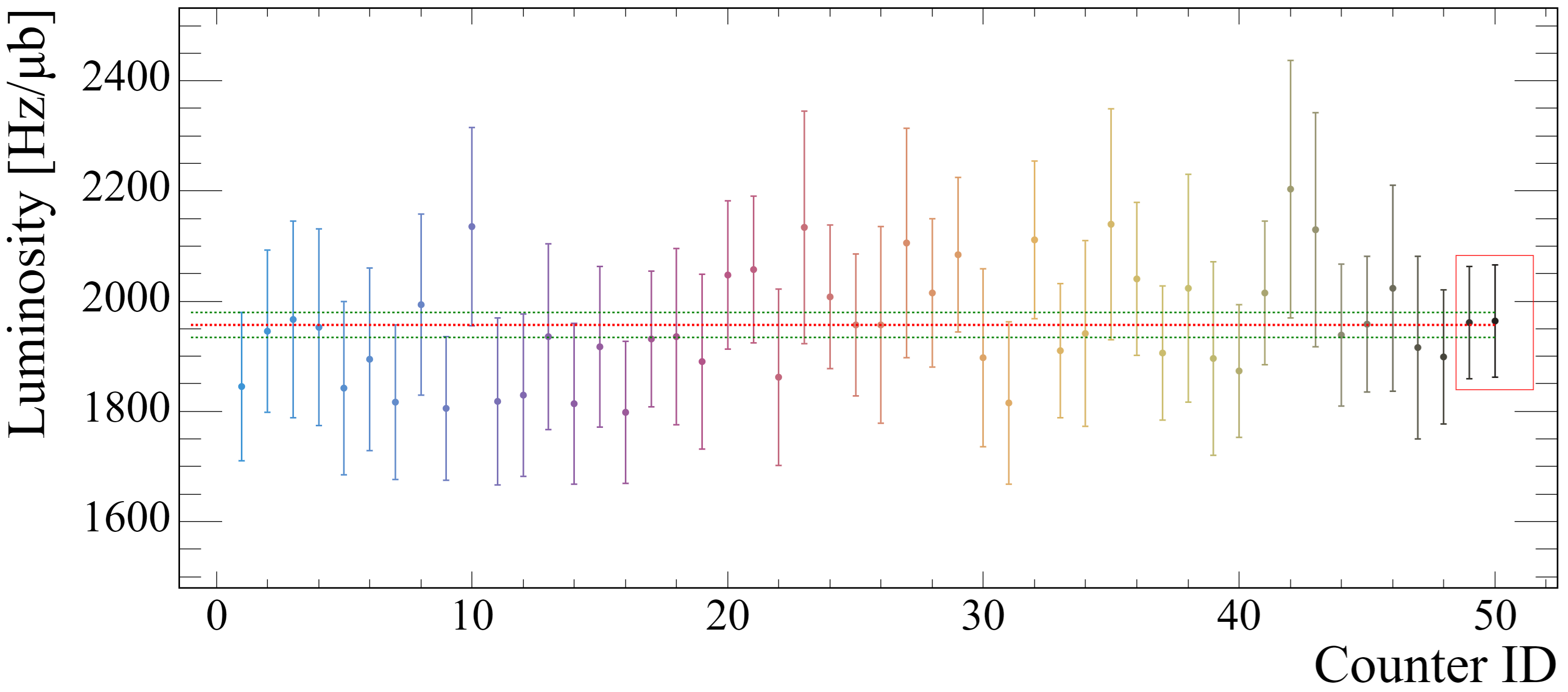}
    \caption{Luminosity is estimated from a subset of 48 hit counters, equally divided between \richone (leftmost) and \richtwo (rightmost). Error bars are statistical only. The last two values, highlighted in the red box, are obtained using the number of \velo tracks as luminosity counters. The average luminosity value obtained using the \rich counters is shown as a dotted horizontal line. The interval corresponding to one standard deviation is also indicated with green lines.}
    \label{fig:luminominal}
\end{figure}

For each PDM, the average of the cross-section values measured in each colliding BXID is employed to determine the luminosity for a given running period using Eq.~\ref{eq:lumiLHCb}, that for a generic luminosity proxy and as a generalisation of Eq.~\ref{eq:lumiLHCb}, becomes

\begin{equation}
\label{eq:lumi_formula}
    \mathcal{L} = {\frac{\mu_{\text {vis}}}{\sigma_{\text {vis}}}} N_\text{bb} f_r,
\end{equation}

\noindent where $\mu_{\mathrm{vis}}$ is measured through the log-zero method in the time interval of interest during physics data-taking. In order to measure $\mu_{\mathrm{vis}}$ during full machine operations, the $be$ and $eb$ background has to be evaluated and subtracted. The scaling of the luminosity given by the \rich hit counters is studied in a dedicated luminosity scan where the instantaneous luminosity is changed in known steps up to the nominal \lhcb luminosity. The value of $\mu_{\mathrm{vis}}$ is estimated by considering only the leading bunch in batches of consecutive colliding bunches in \lhcb. This choice is made because, for these bunches, the background is under control, and effects such as spillover and residual material activation do not have a significant impact. The systematic uncertainty associated with this choice is estimated by including trailing bunches. As shown in Fig.~\ref{fig:luminominal}, the values obtained from the hit counters using Eq.~\ref{eq:lumi_formula} are compared with the luminosity estimated using the number of tracks detected in the \lhcb vertex detector (\velo), which has been established as the most stable \lhcb luminosity counter during \lhc Run 1 and 2~\cite{LHCb-PAPER-2014-047}, showing excellent consistency. The measured instantaneous luminosity from the \rich in a run acquired at nominal pile-up, averaging the values from the subset of 48 hit counters, is

\begin{equation}
\label{eq:lumiResults}
    \lum_\text{RICH} = 1938 \pm 23~(\mathrm{stat}) \pm 97~(\mathrm{syst_\mathrm{vdM}}) \pm 78~(\mathrm{syst_{bkg}})~\hz/\mub,
\end{equation}

The first uncertainty is statistical and the systematic uncertainty is divided in two contributions. The former, syst$_{\mathrm{vdM}}$, corresponds to a conservative estimate of the systematic error associated to the cross-section measurement in the vdM scan, including also the missing corrections to be applied, such as ghost and satellites charge, length scale calibration, beam-beam deflection and efficiency of the observable. A 5\% contribution has been assigned as syst$_{\mathrm{vdM}}$, based on the study performed in Ref.~\cite{LHCb-PAPER-2014-047}. This contribution will be significantly reduced once a comprehensive offline luminosity analysis will be performed with Run 3 data. The other contribution to the systematic uncertainty, syst$_{\mathrm{bkg}}$, is related to the $\mu_{\text {vis}}$ evaluation, appearing as the numerator in Eq.~\ref{eq:lumi_formula}. The background subtraction in Eq.~\ref{eq:muvis} need to be evaluated in physics data-taking and can affect the luminosity value. The value of syst$_{\mathrm{bkg}}$ is assigned by estimating the background using colliding BXIDs in different positions inside the filling scheme, as in isolated $ee$ BXIDs or after a batch of colliding bunches. The value of $\mu_{\mathrm{vis}}$ is also evaluated using leading or trailing colliding bunches in a batch. This systematic contribution will also be reduced in a precise luminosity measurement, after the assessment of the background entering each \lhcb counter. 

\section{Calibration of real-time monitors}
\label{sec:currents}
\begin{figure}
    \centering
    \includegraphics[width=0.7\linewidth]{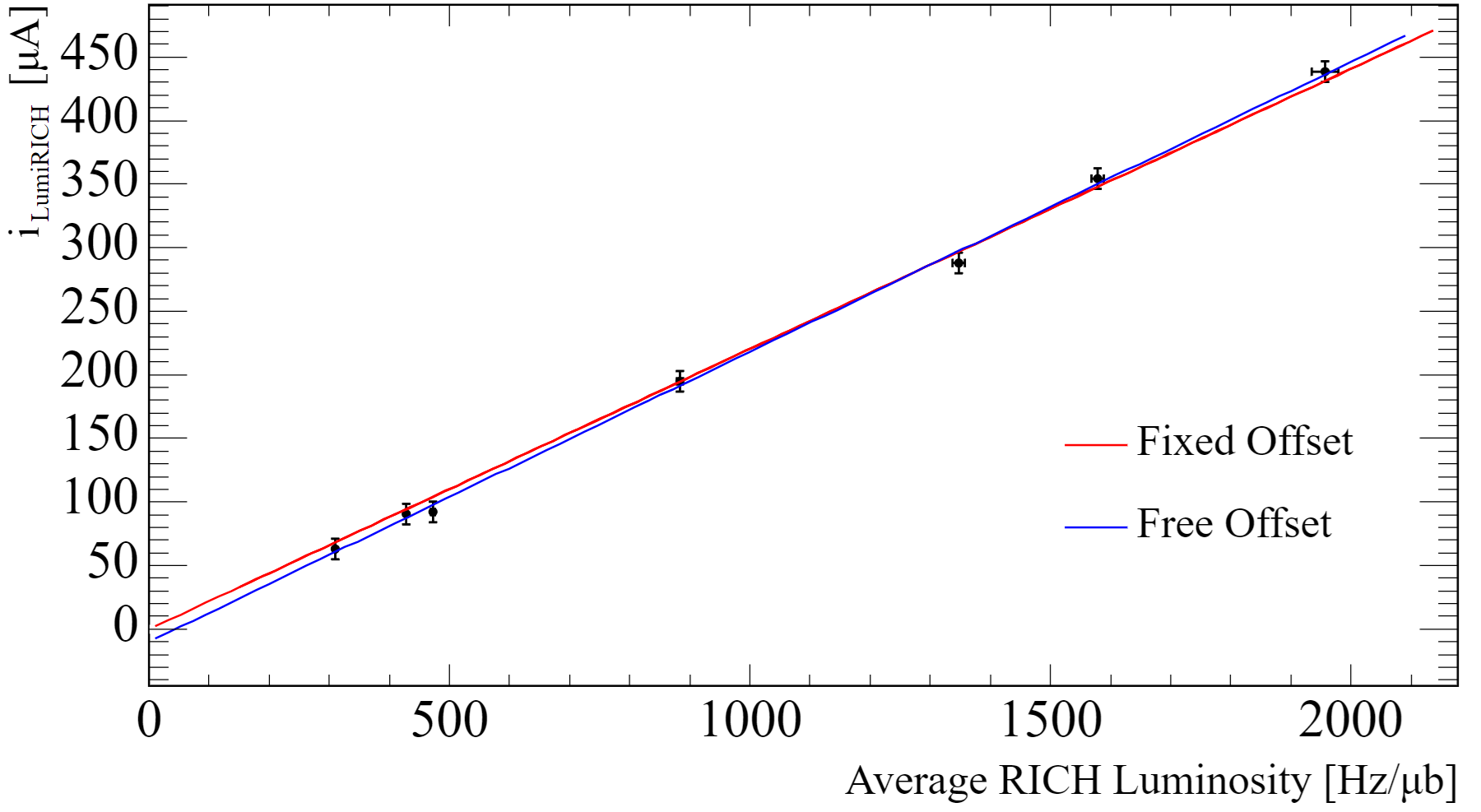}
    \caption{Calibration of an anode current read out from a PDM. The anode current proxy $i_{\mathrm{LumiRICH}}$ value for a given PMT is plotted against the luminosity measured using the RICH hit counters. Each point corresponds to a step of a luminosity scan which covered the full luminosity range of LHCb, up to the nominal Run 3 value. The fit is performed with a free offset (fitted slope = $0.228\pm0.005\;\mu$A$\;\mu$b\;s and offset = $-10\pm6\;\mu $A) or fixing the offset to zero (fitted slope = $0.220\pm0.003\;\mu$A$\;\mu $b\;s), with a difference in the estimated luminosity of less than 1\%, used as a systematic error in the calibration. The procedure is repeated for all PDMs in the \richone high-occupancy region.}
    \label{fig:currentCalibration}
\end{figure}

\begin{figure}[tb]
  \begin{center}
    \includegraphics[width=0.49\linewidth]{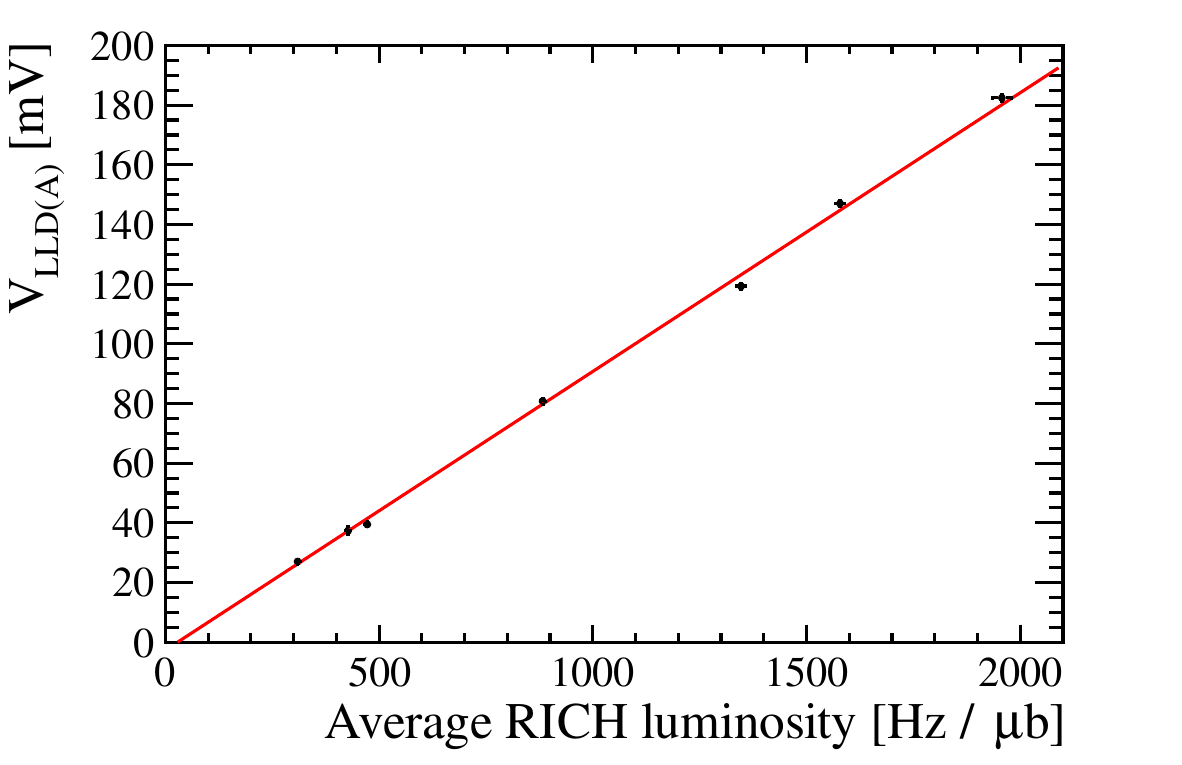}\put(-180,120){(a)}
    \includegraphics[width=0.49\linewidth]{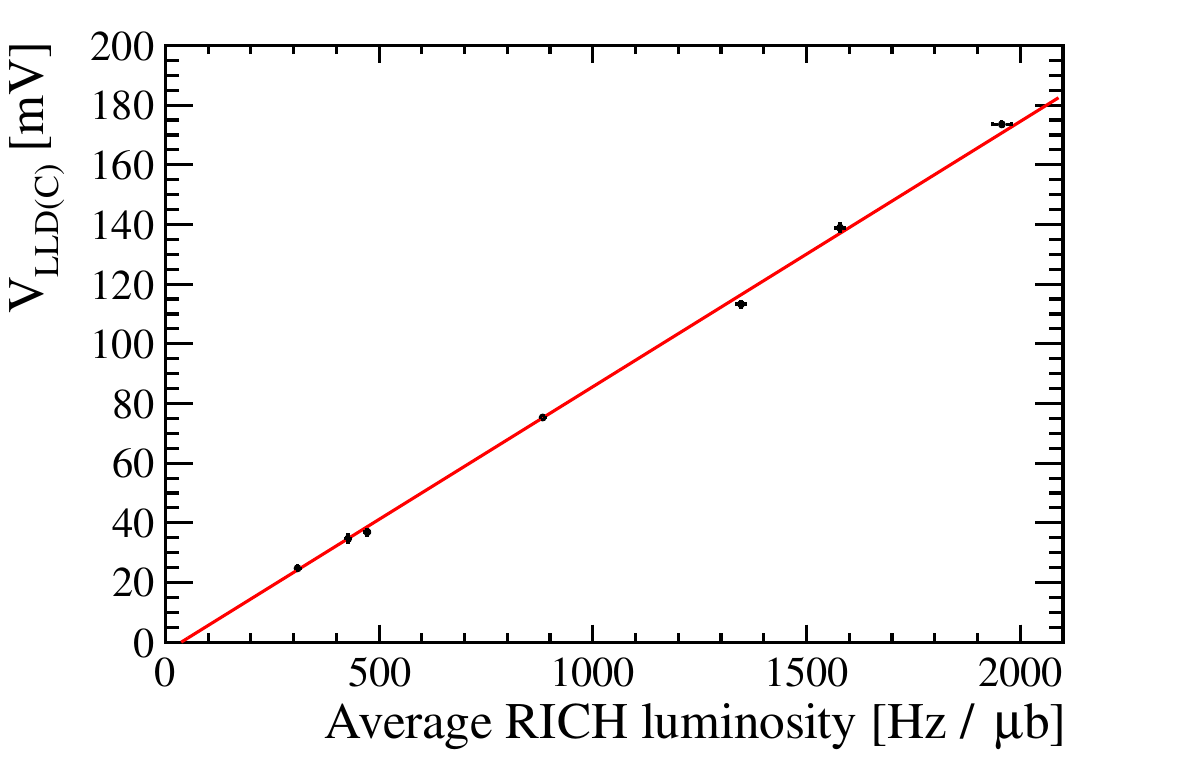}\put(-180,120){(b)}
    \vspace*{-0.5cm}
  \end{center}
  \caption{Correlation between the average \rich luminosity and the LLD signals for \richtwo (a) A-side and (b) C-side (right). The baseline values are subtracted before the linear fit is performed. For the A-side the offset is $-2.7\pm0.6$ mV and the slope is $0.093\pm 0.001$ mV$\;\mu$b\;s, the C-side fitted values are $-3.3\pm0.6$ mV and $0.089\pm0.001$ mV$\;\mu$b\;s for offset and slope, respectively.}
  \label{fig:LLDVsiAverageLumiRich}
\end{figure}

The real-time luminosity determination plays an important role in the operation of \lhcb and provides a prompt feedback during data taking. The relation between the anode currents of the MaPMTs and the luminosity has been extensively discussed in Sec.~\ref{sec:operations}, reporting how a direct calibration of  $i_{\text{LumiRICH}}$ is not possible during vdM scans, since only about 20 bunches collide in \lhcb which is below the minimum rate required to observe a sizeable increase in the power-supply currents. The values of $i_{\text{LumiRICH}}$ is therefore calibrated with the luminosity estimated through the hit counters. The anode currents available from each PDM in the high-occupancy region scales during the luminosity scan as shown in Fig.~\ref{fig:ecsOperations}, and, for each step, the corresponding luminosity is measured using Eq.~\ref{eq:lumi_formula}. The same procedure is used to calibrate the values of $V_\text{LLD}$. Figures~\ref{fig:currentCalibration} and~\ref{fig:LLDVsiAverageLumiRich} show the results of the calibration for the $i_{\text{LumiRICH}}$ and $V_\text{LLD}$ variables, respectively.

\begin{figure}
    \centering
    \includegraphics[width=0.8\linewidth]{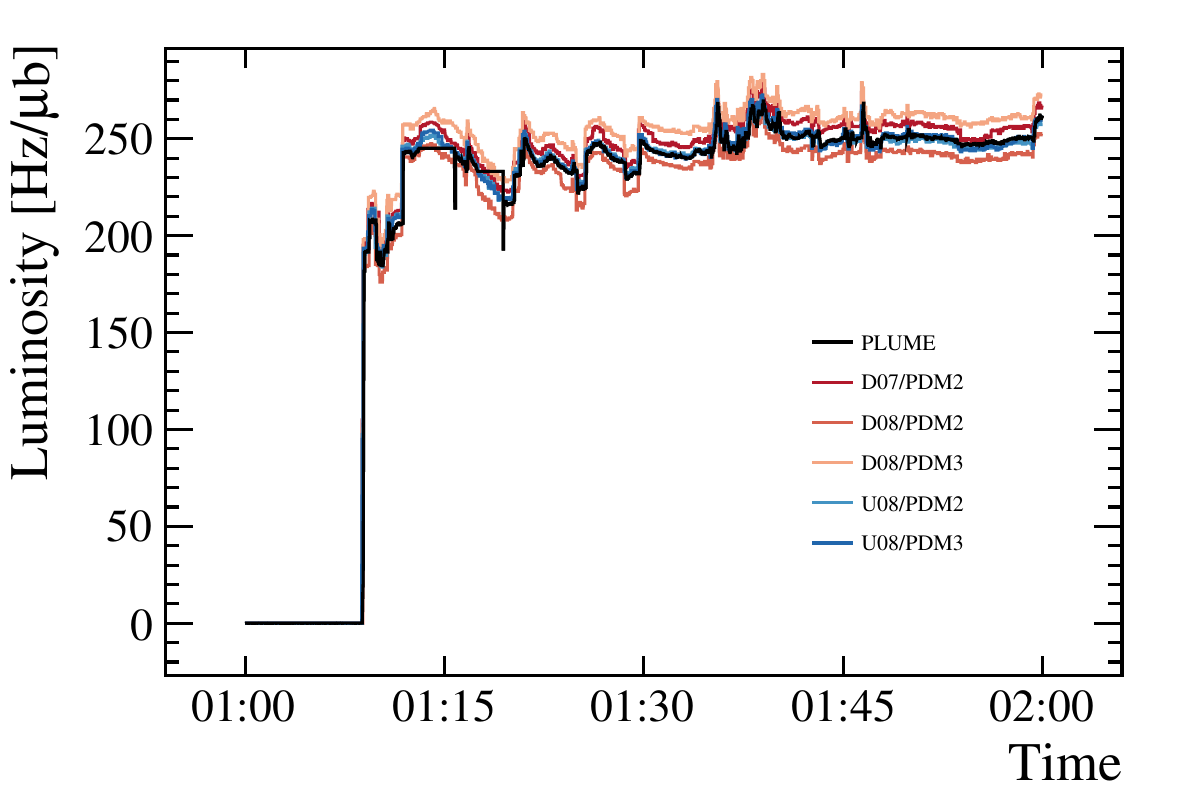}
    \caption{The luminosities determined through the \rich calibrated currents compared to the values provided by Plume as a function of time. The period of time corresponds to the start of fill, where the variations are due to the levelling procedure and beam adjustments, as opposed to noise in the detectors. Agreements is excellent, and the plots are available for monitoring purposes in \lhcb.}
    \label{fig:currentVsPlume}
\end{figure}

After the previous calibration, the values of $i_{\text{LumiRICH}}$ are converted into instantaneous luminosity values and published in the online \lhcb monitoring framework. The luminosity estimated from the \rich anode currents are shown in Fig.~\ref{fig:currentVsPlume} together with the value provided by the PLUME detector as a reference. The spread in the luminosity values estimated using the \rich detectors, even with a preliminary calibration neglecting some of the background sources induced by beam-gas interactions, is below the $\pm 5\%$ required by the experiment for luminosity levelling.

\section{Summary and prospects}
\label{sec:results}

The upgraded \rich detectors can be used to provide a measurement of luminosity for the \lhcb experiment. Linearity of the MaPMT anode currents, enabled in \richone by powering the last dynode, allows a real-time determination of the luminosity, an alternative to the one provided by the PLUME detector, which can increase redundancy during \lhcb operations. These real-time monitors are calibrated with the information available from the \rich hit counters, avoiding a continuous cross-calibration with other \lhcb counters. Furthermore, the signal detected by the LLD safety system can be used as an additional real-time monitor, independent of the accelerator or beam modes of the \lhc. The procedure to integrate these signals in the \lhcb control system has been commissioned and will be used as a backup signal to be provided to the \lhc accelerator team. The relative variations across the values of $i_{\text{LumiRICH}}$ is within $\pm 5\%$ when also including a systematic uncertainty related to the subtraction of the baseline, fulfilling the requirements of the \lhcb experiment in terms of luminosity levelling. 

As discussed in Sec.~\ref{sec:VdM}, a measurement of the instantaneous luminosity obtained when using the subset of the hit counters available offline to the \rich system when running at nominal pile-up conditions gives a value

\begin{equation}
    \lum_\text{RICH} = 1938 \pm 23~(\mathrm{stat}) \pm 133~(\mathrm{syst})~\hz/\mub,
\end{equation}

\noindent where the first uncertainty is statistical and the second is a combination of systematic uncertainties, expected to decrease in the future with additional measurements. 

Future prospects involve further improvements in the calibration of the real-time luminosity counters, including the subtraction of beam-gas events and correction factors associated with the \velo detector closing around the interaction point. The former correction is particularly relevant given the concurrent $pp$ and $p$-gas operations in \lhcb, involving the injection of different gases in the SMOG2 system~\cite{LHCb-TDR-020}. This is estimated to give a correction factor at the percent level. The real-time luminosity counters are also affected by the different number of hits due to secondary particles produced upstream during and after the \velo is closed, varying by up to 30\% depending on the region under consideration. Approximately six minutes are required for the \velo to move from the open to the closed position at the beginning of a physics fill: a linear correction will be applied to take into account this data-taking phase. Regarding the LLD-based real-time monitors, an upgrade of the system has been made by installing a new Trans-Impedance Amplifier readout for 2024 data-taking, allowing further improvement to the linearity with luminosity and it is also going to be integrated into the online monitoring.

In conclusion, the feasibility of a luminosity measurement, both online and offline, using the RICH detectors has been demonstrated. The RICH luminosity proxies are fully integrated into the LHCb data flow. Further improvements of the measurement will be achieved through a comprehensive offline \lhcb luminosity analysis.

\section*{Acknowledgements}

\noindent We thank the \lhcb \rich and Luminosity teams for supporting this publication, and in particular Neville Harnew, Clara Matteuzzi and Valeriia Zhovkovska for the review. We express our gratitude to our colleagues in the CERN
accelerator departments for the excellent performance of the LHC. We
thank the technical and administrative staff at the LHCb
institutes.
We acknowledge support from CERN and from the national agencies:
CAPES, CNPq, FAPERJ and FINEP (Brazil); 
MOST and NSFC (China); 
CNRS/IN2P3 (France); 
BMBF, DFG and MPG (Germany); 
INFN (Italy); 
NWO (Netherlands); 
MNiSW and NCN (Poland); 
MCID/IFA (Romania); 
MICINN (Spain); 
SNSF and SER (Switzerland); 
NASU (Ukraine); 
STFC (United Kingdom); 
DOE NP and NSF (USA).
We acknowledge the computing resources that are provided by CERN, IN2P3
(France), KIT and DESY (Germany), INFN (Italy), SURF (Netherlands),
PIC (Spain), GridPP (United Kingdom), 
CSCS (Switzerland), IFIN-HH (Romania), CBPF (Brazil),
and Polish WLCG (Poland).
We are indebted to the communities behind the multiple open-source
software packages on which we depend.
Individual groups or members have received support from
ARC and ARDC (Australia);
Key Research Program of Frontier Sciences of CAS, CAS PIFI, CAS CCEPP, 
Fundamental Research Funds for the Central Universities, 
and Sci. \& Tech. Program of Guangzhou (China);
Minciencias (Colombia);
EPLANET, Marie Sk\l{}odowska-Curie Actions, ERC and NextGenerationEU (European Union);
A*MIDEX, ANR, IPhU and Labex P2IO, and R\'{e}gion Auvergne-Rh\^{o}ne-Alpes (France);
AvH Foundation (Germany);
ICSC (Italy); 
GVA, XuntaGal, GENCAT, Inditex, InTalent and Prog.~Atracci\'on Talento, CM (Spain);
SRC (Sweden);
the Leverhulme Trust, the Royal Society
 and UKRI (United Kingdom).

\addcontentsline{toc}{section}{References}
\bibliographystyle{LHCb}
\bibliography{main,standard,LHCb-PAPER,LHCb-CONF,LHCb-DP,LHCb-TDR}

\end{document}